\documentclass[aps,12pt,a4paper]{revtex4}
\usepackage{graphicx}
\usepackage{amsmath,amssymb,color}
\usepackage[english]{babel}
\usepackage{mathtools}

\parskip=\medskipamount

\newcommand{\eq}[1]{(\ref{#1})}
\newcommand{\fig}[1]{Fig.~\ref{#1}}

\newcommand{\be}{\begin{equation}}
\newcommand{\ee}{\end{equation}}

\newcommand\disp{\displaystyle}

\newcommand{\im}{\textrm{Im}\,}

\begin{document}

\title{Concepts of polymer statistical topology}

\author{Sergei Nechaev \medskip \\
Universit\'e Paris-Sud/CNRS, LPTMS, \\ 91405 Orsay, France \medskip \\
CNRS/Independent University of Moscow, \\  Poncelet Lab., Moscow, Russia \medskip \\
P.N. Lebedev Physical Institute RAS, \\ 119991, Moscow, Russia}


\begin{abstract}

I review few conceptual steps in analytic description of topological interactions, which constitute
the basis of a new interdisciplinary branch in mathematical physics, emerged at the edge of
topology and statistical physics of fluctuating non-phantom rope-like objects. This new branch is
called statistical (or probabilistic) topology.

\end{abstract}

\maketitle

\tableofcontents

\section{Introduction: What we are talking about?}
\label{sect:1}

How to peel an orange, without removing its skin? Can one make an omelet from unbroken eggs? Can
one smoothly (without tops) comb hairy billiard ball (the sphere), or a donut (the torus)? Why one
cannot tie a knot on a telegraph wire, stretched along the railway line? These and similar, often
entertaining and seemingly naive questions are directly related to the topology. I will be
interested in a rather narrow range of problems associated with so-called low-dimensional topology,
i.e., with the topology of systems containing long linear hurly-burly threads of different physical
nature. As these objects, one can play with polymeric chains, vortex lines in superconductors,
strings in quantum field theory, etc.

It is worth noting that the low-dimensional topology is a pretty insidious, or, better to say
"serpentine" science, because the daily use of ropes and wires, being trivial for "kitchen tasks",
becomes almost useless for mathematical description of knots. Careful tying shoe laces does not
help much in constructing topological invariants, determination of physical and geometric
properties of highly entangled polymeric networks, folded proteins, DNAs in chromosomes, etc.
Moreover, even obvious notions form the everyday's experience seem to be incorrect: for example,
any child knows what the knotted rope is, and what is its difference from the unknotted one,
however, this knowledge fails being translated into mathematically rigorous terms: one cannot
define a knot on any open curve, and to talk seriously about knots, only closed paths should be
dealt with. Indeed, having free ends of the thread, one can always transform any two arbitrary
conformations of threads into each other by a continuous deformation. Therefore, for open threads
it would be more correct to speak of quasi-knots (instead of knots) tied on a closed curve
consisting of the thread itself, and, say, the segment joining its open extremities. However, in
the case when the knot has a size substantially smaller than the thread's length, the difference
between the knot and quasi-knot becomes negligible.

\begin{figure}[htbp]
\centerline{\includegraphics[width=10cm]{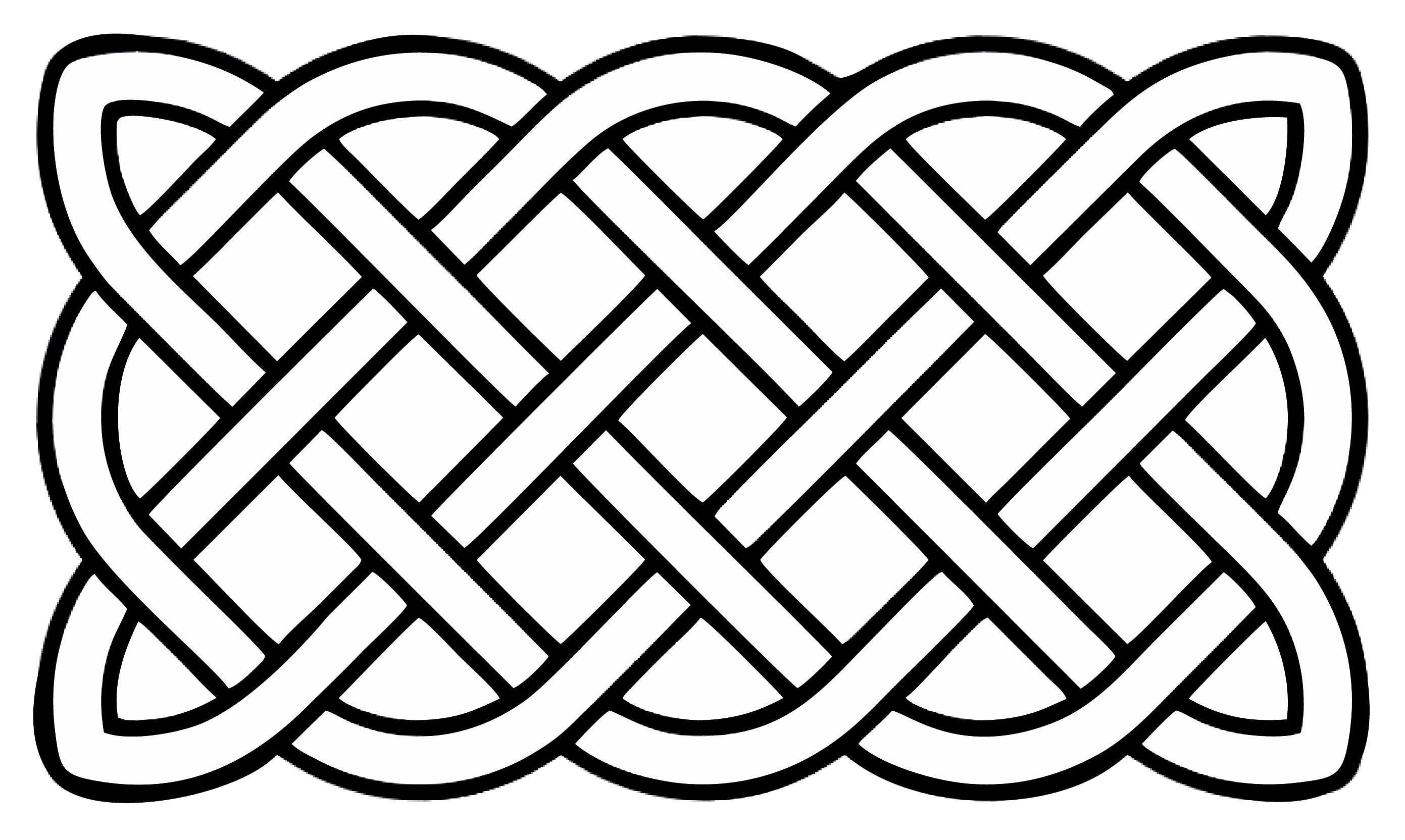}}
\caption{Sample of celtic knot projection to the plane (knot diagram).}
\label{fig:01}
\end{figure}

History has brought to us the name of one of the first topologists-experimenters, Alexander the
Great (IV century BC), whose being unable to untie the Gordian knot, just slashed it with the
sword. Surprisingly, modern algebraic topology partly borrows ideas of Alexander the Great to build
topological invariants by splitting intersections of wires on a plane knot projection. For
introduction to these constructions, please look \cite{kauffman}. Nobody knows how the Gordian knot
looked exactly, however there is an opinion that it resembled so-called celtic knot, whose typical
plane projection (knot diagram) is shown in the Fig.\ref{fig:01}.

Unknotting heavily tangled ropes, we are not even surprised that long threads, left to themselves,
have the feature of entangled most unpleasant way of all. We are just a bitter sigh, when almost
unravelled ball of wool accidentally falls into the clutches of a curious kitten, and our hard work
goes to hell...

As a rule, we do not think about physical reasons for such "injustice", considering it as one of
manifestations of the general law of life: unpleasant things happen more often than pleasant ones,
and the sandwich usually falls butter side down. And we are quite aware that the spontaneous
knotting of long strands, managed by the laws of probability theory, is a consequence of
non-Euclidean geometry of the phase space of knots, i.e., such hypothetical space that contains all
possible knots and in which there is a concept of the metric, or the distance between
"topologically similar" knots. The more different topological types of two knots, the greater the
distance between them in this space. In general, the construction of this space, called the
universal covering, and the description of its properties is extremely challenging. However, for
some special cases, to which I address a little later, this space can be exhaustively described in
geometric terms. So, it turns out that, in order to untie ropes in a purely scientific way, one has
first to understand the Lobachevsky-Riemann geometry and its relation to the knot theory, and then
to learn how the probability theory works in this non-Euclidean space. Before going further, let me
briefly digress and discuss some of physical questions in which the topological interactions play a
crucial role.

\begin{figure}[htbp]
\centerline{\includegraphics[width=12cm]{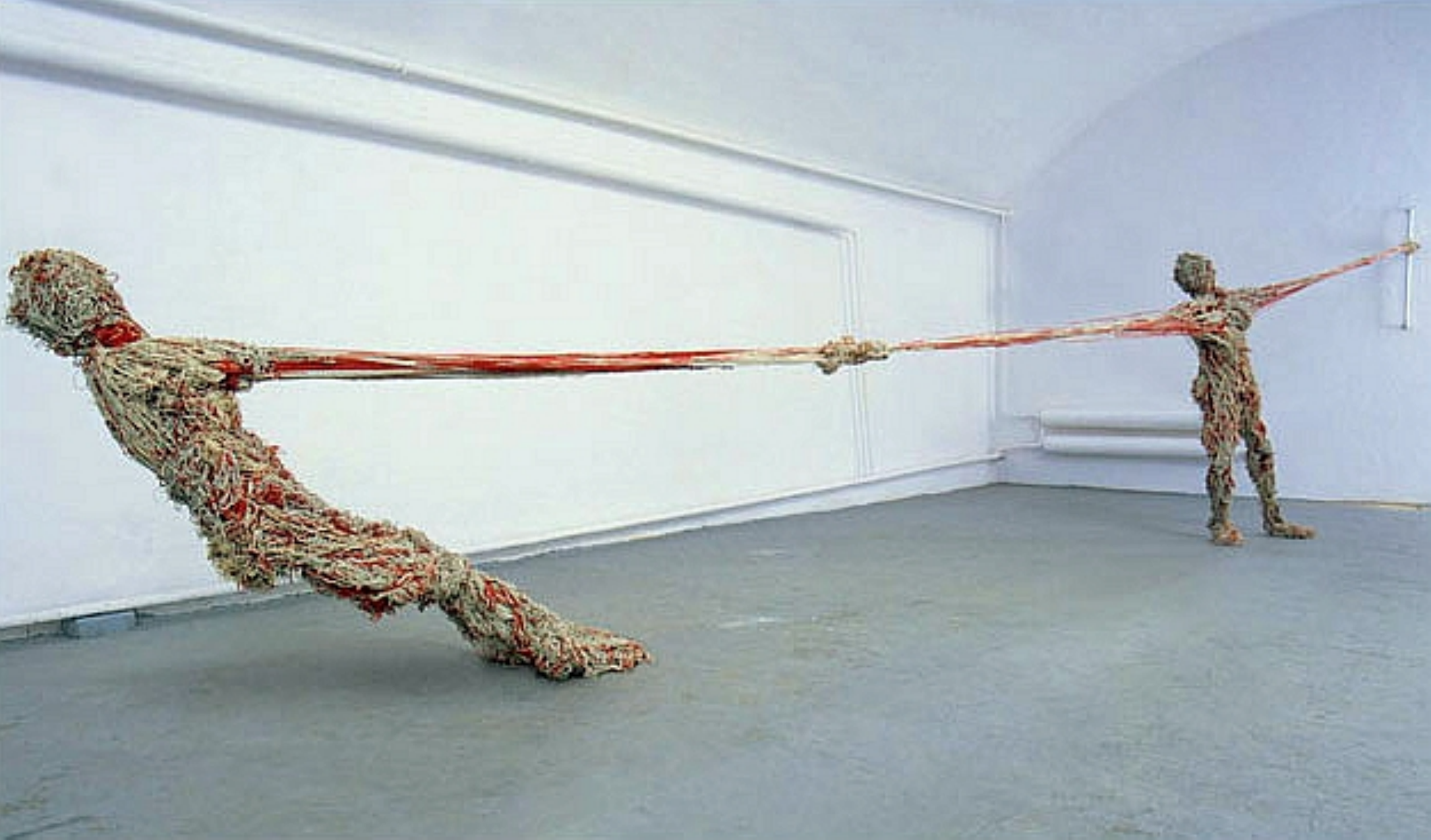}}
\caption{Emotional view on elastic deformation. This sculpture, displayed at the Ujazdowski
Castle's Center for Contemporary Art in Warsaw (Poland), was made by Agnieszka Kalinowska in 2002.}
\label{fig:02}
\end{figure}

Mathematicians, primarily are interested in the question how to construct characteristics of
entangled curves, that depend only on their topological state, but not on their shapes. Besides the
traditional fundamental topological issues concerning the construction of new knot invariants,
description of knot homologies, homotopic classes etc, there exists an important set of adjoint but
much less studied problems related to probability theory and statistical physics. First of all, I
mean the problem of so-called "knot entropy" evaluation. Suppose we know everything about a set of
entangled curves and know how to classify their topology. In order to understand which issue is
interesting for physicists, imagine that one has a sample of rubber, i.e. the collection of polymer
chains crosslinked at their extremities such that they form a connected network. When the network
is deformed as shown in the Fig.\ref{fig:02}, the ensemble of available conformations of each
individual subchain in the network changes, resulting in an essential decrease of the entropy, $S$.
So, one says, that the elasticity of stretched rubber is of entropic nature.

Once being prepared, the network sample cannot change its topology, i.e. is \emph{topologically
frozen}. If the applied deformations do not break the subchains, the topological constraints like
entanglements between different subchains, at high extensions enter into the game as new effective
quasi-links, providing additional restrictions on the ensemble of available subchain conformations.
Therefore, the topological constraints contribute to the stress-strain dependence of polymeric
networks, being the origin of so-called Mooney-Rivlin corrections to the classical Hook's law
\cite{grosberg-khokhlov}.

In general, the non-phantomness of a polymer chain causes two types of interactions: i) volume
interactions vanishing for infinitely thin chains, and ii) topological interactions, which survive
even for chains of zero thickness. For sufficiently high temperatures, a polymer molecule strongly
fluctuates without a reliable thermodynamic state called a coil state. However for temperatures
below some critical value, $\theta$, the polymeric chain forms a weakly fluctuating dense globular
(drop-like) structure \cite{grosberg-khokhlov}, and one may expect that just in the globular state
the topological interaction manifest themselves in all their glory. The crucial difficulty in
description of topological interactions comes from their non-locality: the entropic part of a
polymeric chain free energy, $F=-TS$, strongly depends on the global chain topology. Saying more
formal, the topological interactions in dense polymer systems cannot be treated in a
\emph{perturbative} way and new ideas of \emph{nonperturbative} description are demanded.

The general problem we are dealing with, can be formulated as follows. Consider a 3D cubic grid,
and let $\Omega$ be the ensemble of all possible closed non-self-intersecting $N$-step paths on
this grid with one point fixed. Denote by $\omega$ ($\omega\in \Omega$) some particular realization
a path. We are aimed to calculate the partition function, $Z\{{\rm Inv}\}$, for a knot to be in a
specific topological state characterized by the topological invariant ${\rm Inv}$ (yet
non-specified). This can be written formally as
\be
Z\{{\rm Inv}\}=\sum \limits_{\{\omega \in \Omega\}} \Delta \Big\{{\rm Inv}(\omega)-{\rm Inv}\Big\}
\label{eq:01}
\ee
where ${\rm Inv}(\omega)$ is a functional representation of the invariant for the path $\omega$,
$\rm Inv$ is a specific topological invariant of the knot which we would extract, and $\Delta$ is
the Kronecker delta-function. The entropy, $S\{{\rm Inv}\}$, of a topological state, ${\rm Inv}$,
is defined as
\be
S\{{\rm Inv}\}=\ln Z\{{\rm Inv}\}
\label{eq:02}
\ee
Based on the above definition, we can see that the statistical-topological problems are similar to
those encountered in physics of disordered systems and in particular, of spin glasses. Indeed, the
topological state of a path plays the role of a "quenched disorder" and the entropy, $S\{{\rm
Inv}\}$, is averaged over the ensemble of trajectories fluctuating at the "quenched topological
state" \cite{ne-gr1}. In the context of this analogy, it seems challenging to extend the concepts
and methods developed over the years for disordered systems to the scopes of statistical topology.
The main difference between the systems with topological disorder and the standard spin systems
with disorder in the coupling constant, deals with the strongly nonlocal character of interactions
in the first case: a topological state is determined for the entire closed path and is its "global"
property.

Below I review few conceptual steps in analytic description of topological interactions, which
constitute the basis of a new interdisciplinary branch in mathematical physics, emerged at the edge
of topology and statistical physics of fluctuating non-phantom rope-like objects. This new branch
is called statistical (or probabilistic) topology. Yet its most fascinating manifestation is
connected with the nonperturbative description of DNA packing in chromosomes in a form of a
crumpled globule \cite{gns,gr-rab}. After experimental works of the MIT-Harvard team in 2009
\cite{mirny}, the concept of crumpled globule became a kind of a new paradigm allowing to
understand the mathematical origin of many puzzled features of DNA structuring and functioning in a
human genome.

To intrigue the reader, I can say that the mathematical background of the crumpled globule deals
with the statistics of Brownian bridges in the non-Euclidean space of constant negative curvature.
Forthcoming sections are written to uncover this abracadabra.

\section{Milestones}
\label{sect:2}

\subsection{Abelian epoch}

In 1967 S.F. Edwards laid the foundation of the statistical theory of topological interactions in
polymer physics. In \cite{edwards} he proposed the way of exact computation of the partition
function of a single self-intersecting random walk topologically interacting with the infinitely
long uncrossible string (in 3D), or obstacle (in 2D). Sir Sam Edwards was the first to recognize
the deep analogy between Abelian topological problems in statistical mechanics of Markov chains and
quantum-mechanical problems (like Bohm-Aharonov ones) of the charged particles in the magnetic
field. The review of classical results is given in physical context in \cite{wiegel}, some rigorous
results, including application in financial mathematics were  discussed in \cite{yor}, and modern
advantages are summarized in \cite{grosberg-frisch}. The works of S.F. Edwards opened the "Abelian"
epoch in the statistical theory of topological interactions.

In his work S.F. Edwards used the path integral formalism combined with the functional
representation of the Gaussian linking number. All these steps have been many times reproduced in
the literature, so we do not discuss the details, just remind that one finally arrives at the
quantum problem of a free charged particle (with an imaginary magnetic charge) in a solenoidal
magnetic field. If the magnetic flux (the obstacle) is located at the origin and is orthogonal to
the plane, then for the probability $P(r_1,r_N, \theta,n,N)$ to find a $N$-step polymer chain,
whose extremities are located at the distances $r_1$ and $r_N$ with $\widehat{{\bf r}_1, {\bf r}_N}
= \theta$ and which made $n$ full turns around the origin, we get:
\be
P(r_1,r_N, \theta,n,N) = \frac{e^{-\frac{r_1^2+r_N^2}{2Na^2}}}{\pi Na^2}\int_{-\infty}^{\infty}
d\nu I_{|\nu|}\left(\frac{r_1^2+r_N^2}{2Na^2}\right)e^{i \nu (2\pi n + \theta)}
\label{eq:edw0}
\ee
where $I_{|\nu|}(...)$ is the modified Bessel function of order $|\nu|$, and $a$ is the size of the
monomer (the typical step of the random walk). Obviously, the normalization condition is fulfilled,
\be
\sum_{n=-\infty}^{\infty} P(r_1,r_N, \theta,n,N) = \frac{e^{-\frac{({\bf r}_1-{\bf
r}_N)^2}{Na^2}}}{\pi Na^2}
\label{eq:norm}
\ee
which means that the summation over all windings, $n$, $(-\infty<n<\infty)$ gives the Gaussian
distribution. I will reproduce the result \eq{eq:edw} in the next Section using the conformal
approach. Though it is less popular in polymer physics than the path-integral formalism, it can be
straightforwardly generalized to the non-Abelian multi-obstacle case.

It should be noted that the exact computation of the partition function of the self-intersecting
random walk topologically entangled with \emph{two} uncrossible obstacles in the plane, despite the
huge number of works since 1967, still is an open problem. However, apparently this gap will be
filled soon, because its quantum-mechanical counterpart, the Abelian problem of Bohm-Aharonov
scattering in presence of \emph{two} magnetic fluxes in the plane, has been solved in 2015 by E.
Bogomolny \cite{bogom}. To my point of view, his solution has opened the Pandora's Box in the
field, since he showed the deep mathematical connection of this particular problem to the theory of
Painleve equations, integrable systems etc.

\subsection{Non-Abelian epoch}

\begin{figure}[htbp]
\centerline{\includegraphics[width=10cm]{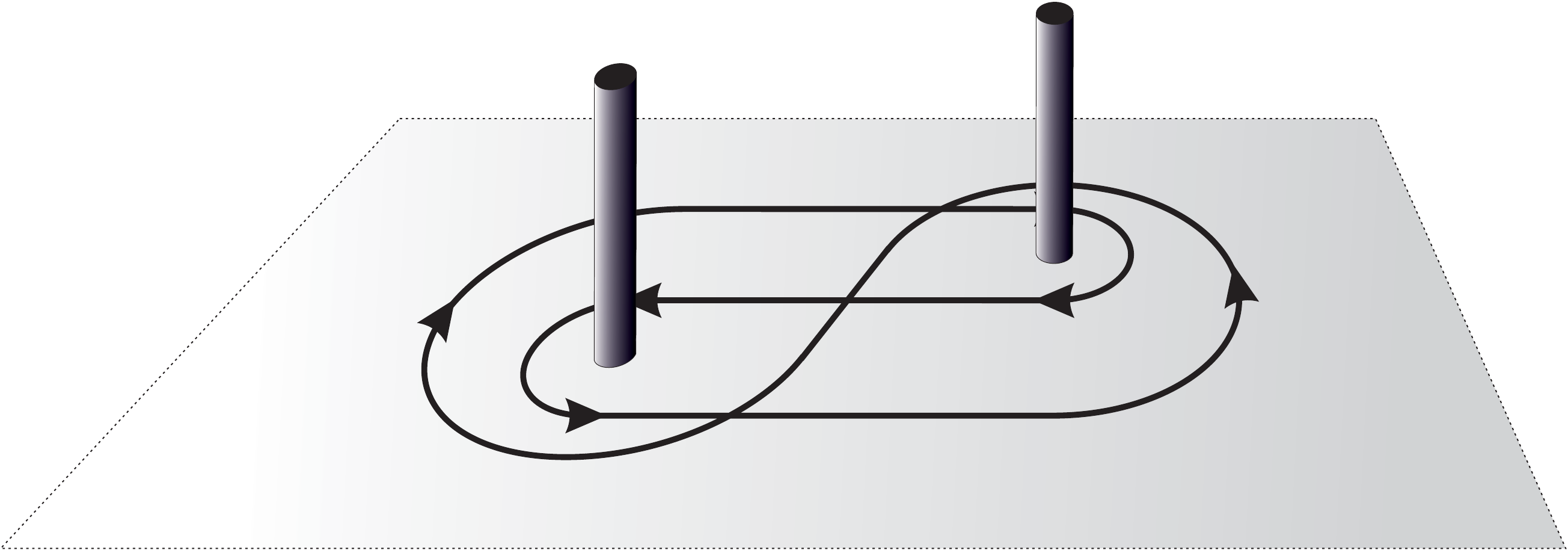}}
\caption{Topological configuration entangled with two obstacles simultaneously but not entangled
with them separately (so-called Pochhammer contour).}
\label{fig:03}
\end{figure}

Each time when we consider statistics of sufficiently dense polymer system, we encounter the
extremely difficult problem of classification of topological states of polymer chains. Even the
simplest physically relevant questions dealing with the knotting probability of a polymer chains,
cannot be answered using the Gauss invariant due to its weakness. The Gauss linking number, becomes
inapplicable since it does not reflect the sequence in which a given topological state was formed.
For example, when some trial trajectory encloses two obstacles, the path is entangled with two
obstacles simultaneously, while being not entangled with them separately, as shown in the
Fig.\ref{fig:03}.

Interestingly, I have not found in the literature any example of the path entangled with
\emph{three} obstacles simultaneously, while not entangled with any separate obstacle and any pairs
of obstacles. Formally the question can be formulated as follows: find the element $X\in F_3$ of a
free group of three generators, $F_3$, such that $X$ belongs both to the commutant of $F_3$ and to
the commutant of commutant of $F_3$.

For dense polymer systems, one encounters many configurations as shown in the Fig.\ref{fig:03}, and
they should be properly classified and treated using the invariants stronger than the Gauss linking
number. So, to summarize, the Abelian (commutative) invariants become inapplicable for dense
polymers and should be replaced by the non-Abelian (noncommutative) ones.

\subsubsection{Methods of algebraic topology in polymer statistics. Topology as quenched disorder}

A very useful and powerful method of knots classification has been offered by a polynomial
invariant introduced by Alexander in 1928. The breakthrough in the field of polymer statistics was
made in 1975-1976 when the algebraic polynomials were used for the topological state identification
of closed random walks generated by the Monte-Carlo method \cite{volog}. It has been recognized
that the Alexander polynomials being much stronger invariants than the Gauss linking number, could
serve as a convenient tool for the calculation of the thermodynamic properties of entangled random
walks. To my point of view, 1975 is the year of the second birth of the probabilistic polymer
topology, since the main part of our modern knowledge on knots and links statistics in dense
polymer systems is obtained with the help of these works and their subsequent developments.

Other polynomial invariants for knots and links were suggested by V.F.R. Jones \cite{Jones} (Jones
polynomials) and by J. Hoste, A. Ocneanu, K. Millett, P.J. Freyd, W.B.R. Lickorish, and D.N. Yetter
\cite{homfly} (HOMFLY polynomials). The Jones invariant arise from the investigation of the
topological properties of braids \cite{Birman}. V.F.R. Jones succeeded in finding a profound
connection between the braid group relations and the Yang-Baxter equations representing a necessary
condition for commutativity of the transfer matrix offered the relation with integrable systems
\cite{collect}. It should be noted that neither the Alexander, Jones, and HOMFLY invariants, nor
their various generalizations are complete, however these invariants are successfully used to solve
many statistical problems in polymer physics. A clear geometric meaning of Jones invariant was
provided by the works of Kauffman, who demonstrated that Jones invariant can be rewritten in terms
of the partition function of the Potts spin model \cite{kauffman}. Later Kauffman and Saleur showed
that the Alexander invariants are related to a partition function of the free fermion model
\cite{Kauf_Sal}. The list of knot invariants used in polymer physics would be incomplete without
mentioning Vassiliev invariants \cite{vassiliev} and Khovanov homologies \cite{khovanov}.

\subsubsection{How to define the knot complexity?}

There are many definitions of knot complexity. Some authors use the concept of  {\it minimal number
of crossings} \cite{tait,bangert,whit1,quake,sheng}. In  other works (see, for example
\cite{whit1,whit2}) knot complexity is  associated with a properly normalized logarithm of a kind
of knot torsion,  $\log |\Delta_K(-1)|$, where $\Delta_K(t)$ is the Alexander polynomial of the
knot  $K$. The estimate of knot complexity using the {\it knot energy} was discussed in the works
\cite{freedman,rolfsen,sossinsky} and to my point of view is yet underappreciated concept by
polymer physicists.

\begin{figure}[htbp]
\centerline{\includegraphics[width=12cm]{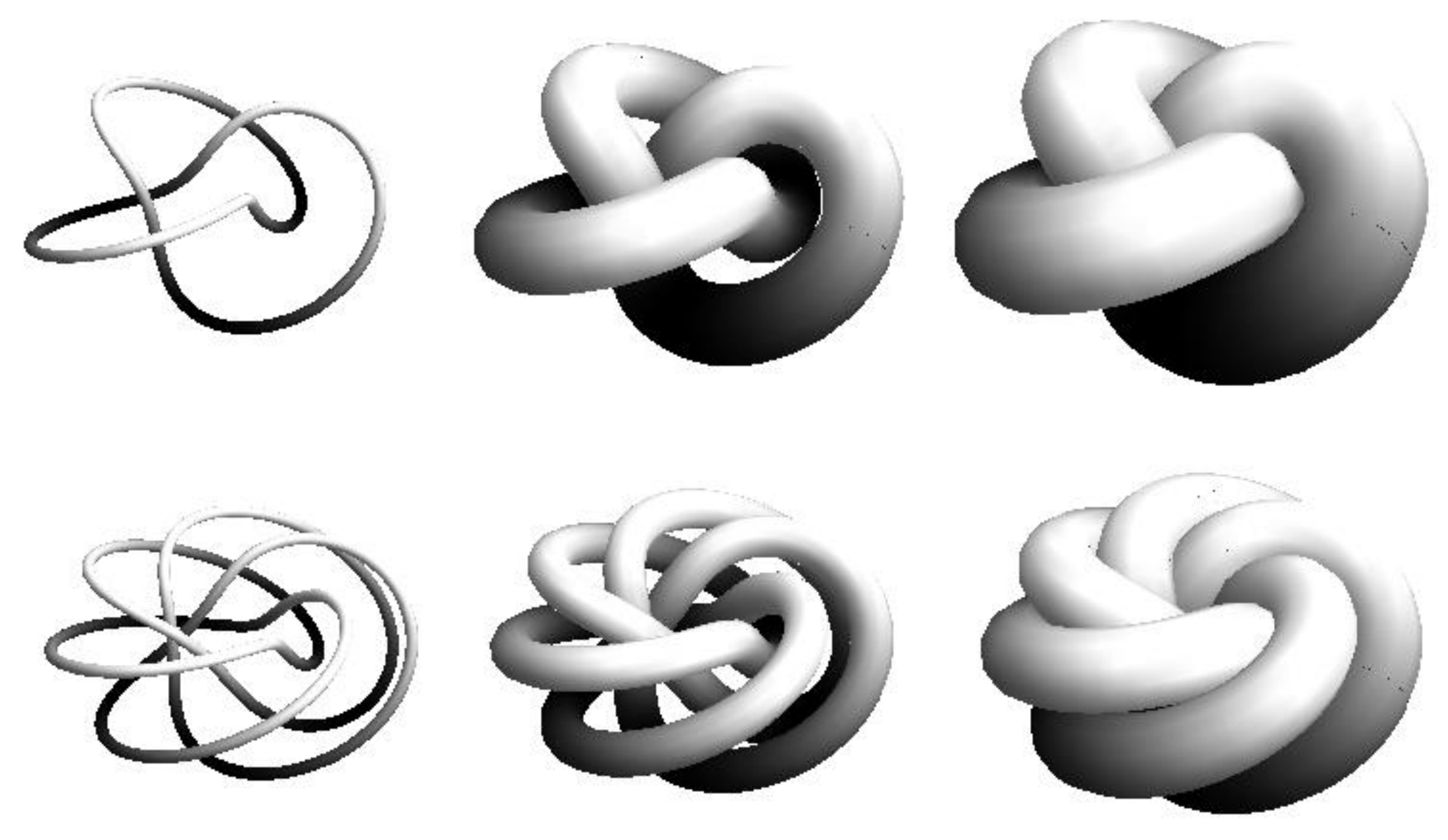}}
\caption{Affine inflation of a torus knot.}
\label{fig:04}
\end{figure}

Another approach deals with the fashionable concept of \emph{knot inflation} \cite{id_kn}. This
topological invariant is defined as the quotient, $\mu$, of the contour length of the knot made of
an elastic tube to its diameter in the maximal uniformly inflated configuration as shown in the
Fig.\ref{fig:04} for two different torus knots of same tube length. One sees that as more complex
the knot, as thinner the limiting tube. Such approach was introduced and exploited in
\cite{gr-rab,id_kn}.

Knot invariants like the minimal number of crossings, as well as those built on the basis of the
knot inflation concept, are similar to the invariants defined as the degree of algebraic polynomial
used in the works \cite{grne_alg2,vasne}. All of them have one common ancestor -- the so-called
\emph{primitive path}, appeared in physical literature in 1970s in the works on entanglements in
polymer melts. Introduced by P. de Gennes (see, for example \cite{gennes}), the primitive path was
served to describe topological effects in the dynamics of individual chains in concentrated polymer
solutions. Later on, the same concept has been used in the computation of equilibrium properties of
polymer chains in lattices of topological obstacles \cite{ne_kh,rub} (I discuss this issue in
details in the next Section).

\begin{figure}[htbp]
\centerline{\includegraphics[width=10cm]{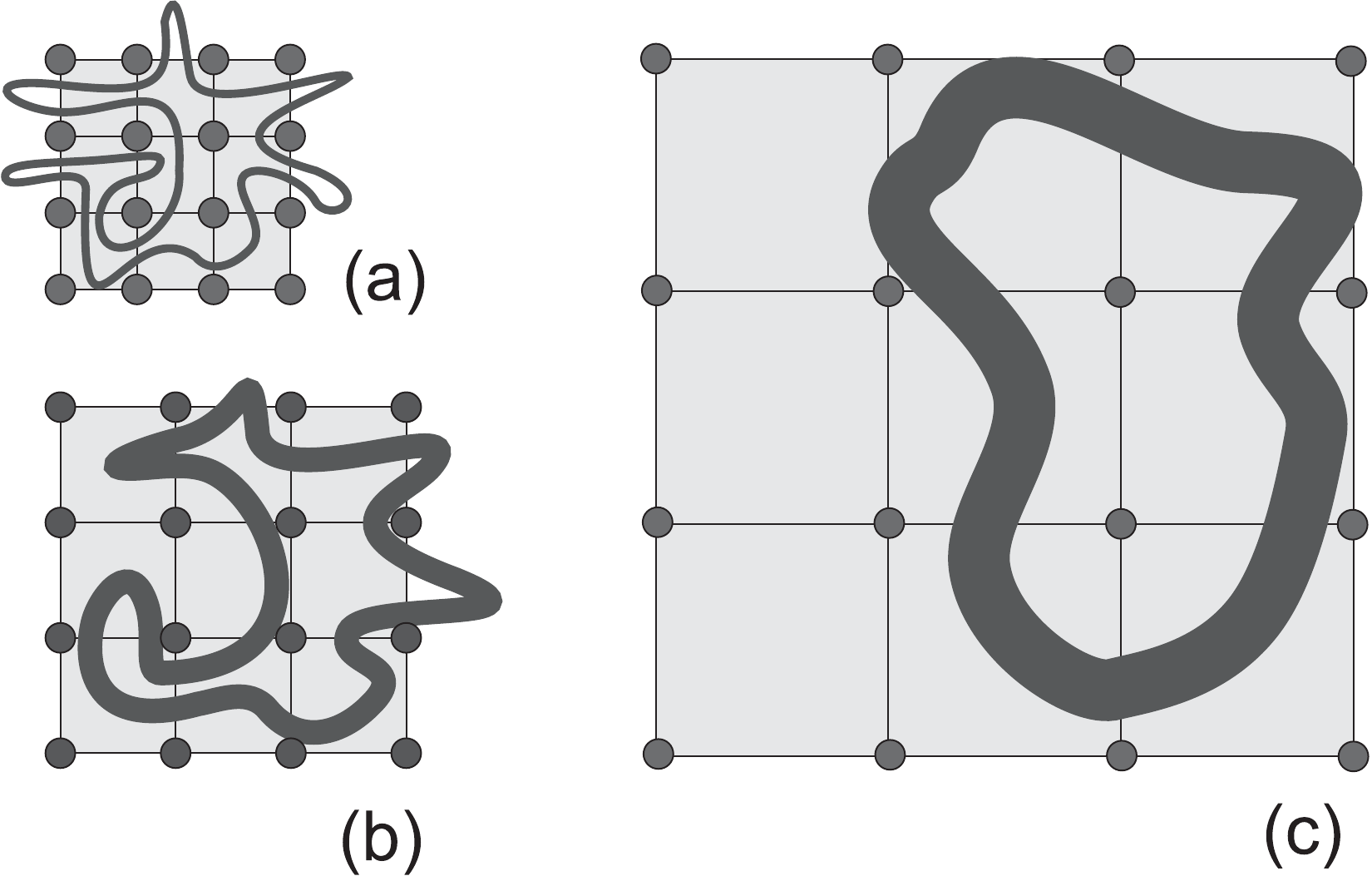}}
\caption{Affine inflation of a lattice of obstacles. The corresponding contour becomes
"less folded".}
\label{fig:05}
\end{figure}

The notion of a primitive path and its relation to the knot inflation concept can be elucidated as
follows. Consider a closed path of fixed length entangled with the lattice of obstacles (see
Fig.\ref{fig:05}a). Performing an affine extension (inflation) of the lattice of obstacles
(preserving the length of the path), one arrives eventually at the unfolded "fully stretched
configuration", see the Figs.\ref{fig:05}b-c. Just the configuration in the Fig.\ref{fig:05}c is
called the \emph{primitive path} and it characterizes the topological state of a path with respect
to the lattice of obstacles. Let us associate the properly normalized length of the chain and the
spacing between obstacles, with the length of an elastic tube in the "knot inflation concept" and
its diameter. In that way, the relationship between the primitive path, the minimal number of
crossings and the quotient $\mu$ in the maximal inflated tube configuration becomes intuitively
clear by construction. In more details this relationship was discussed in \cite{gr-rab}.

In 1991 it was realized \cite{nesin2} that the concept of a primitive path has a straightforward
interpretation in terms of a geodesics in a space of constant negative curvature. In the
forthcoming section, we show how the geodesic length, in turn, may be related via its matrix
representation to the degree of the polynomial invariant. Though our construction is restricted to
the particular case of knots on narrow strips, the very idea can be used to attack more general
models. The relationship between primitive paths and the maximal degree of the corresponding
polynomial invariants has been discussed in \cite{jktr} and will be overviewed in the next Section.

\subsubsection{Conformal methods in statistics of entangled random walks}

In 1985-1988 we have adopted the knowledge of stochastic processes in the Riemannian geometry to
the statistical topology of polymers in multi-punctured spaces \cite{ne_kh,nech-sem,nechJPA}. In
particular, we have shown that the probability for a random walk (a polymer chain without volume
interactions) to be unentangled with the regular lattice of topological obstacles in the plane is
asymptotically described by the probability to form a Brownian bridge (BB) in the Lobachevsky plane
(the non-Euclidean plane with a constant negative curvature). To have a transparent geometric
image, though rather naive, the problem can be formulated as a "snake in a night forest". Suppose
that very long snake, lost in a dense forest at night, would like to grasp randomly its own tale in
such a way, that the formed ring is not entangled with any tree. What is the probability, $P(L,d)$,
of such event, if $L$ is the length of the snake and $d$ is the average distance between the trees?
What is the typical size, $R_g(N,d)$, of the snake (in polymer statistics, where a closed snake is
replaced by a polymer ring, $R_g$ is called the "gyration radius")?

Conformal methods provide straightforward answer to these questions. The key idea is to find the
conformal transformation $w(z)$ which maps the complex plane, $z=x+iy$, with obstacles (branching
points), to the "universal covering" space, $w=u+iv$, free of branchings in any finite domain. The
main ingredient of this approach is the "conformal invariance" of Laplace operator, $\Delta(z)$.
Under the conformal mapping $w(z)$ the Laplacian $\Delta(z) =\partial^2_{xx} +
\partial^2_{yy}$ is transformed to $\Delta(w)=\partial^2_{uu} + \partial^2_{vv}$ as follows
\be
\Delta(w)=\partial^2_{uu} + \partial^2_{vv}= |w'(z)|^{2} (\partial^2_{xx} + \partial^2_{yy})
\label{eq:03}
\ee
where $w'(z)=\frac{dw(z)}{dz}$. Define also $z(w)$, the inverse function of $w(z)$. If we are lucky
enough and found the desired conformal mapping $w(z)$, then, in the universal covering space, our
initial topological problem looks formally extremely simple: we have just to solve the
diffusion-like equation in time $t$ with the diffusion coefficient $D$:
\be
\partial_t P(w,t) - D|z'(w)|^{-2} (\partial^2_{uu} + \partial^2_{vv})P(w,t) =
\delta(w-w_0)\delta(t)
\label{eq:cov}
\ee
without any topological constraints since all information about the topology is encoded now in the
boundary conditions of the corresponding Cauchy problem in the covering space $w$. In the theory of
stochastic processes the equation \eq{eq:cov} describes the diffusion in the "lifted" time, since
it can be considered as a standard diffusion in the new metric-dependent time $T$, where
$\partial_T = |z'(w)|^{2}\partial_t$. However the simplicity of \eq{eq:cov} in majority of cases is
rather illusory: finding conformal mapping $z(w)$, and then solving \eq{eq:cov} analytically, both
these tasks are challenging problems. Despite, few nontrivial cases can be treated and solved at
least asymptotically.

\begin{figure}[htbp]
\centerline{\includegraphics[width=10cm]{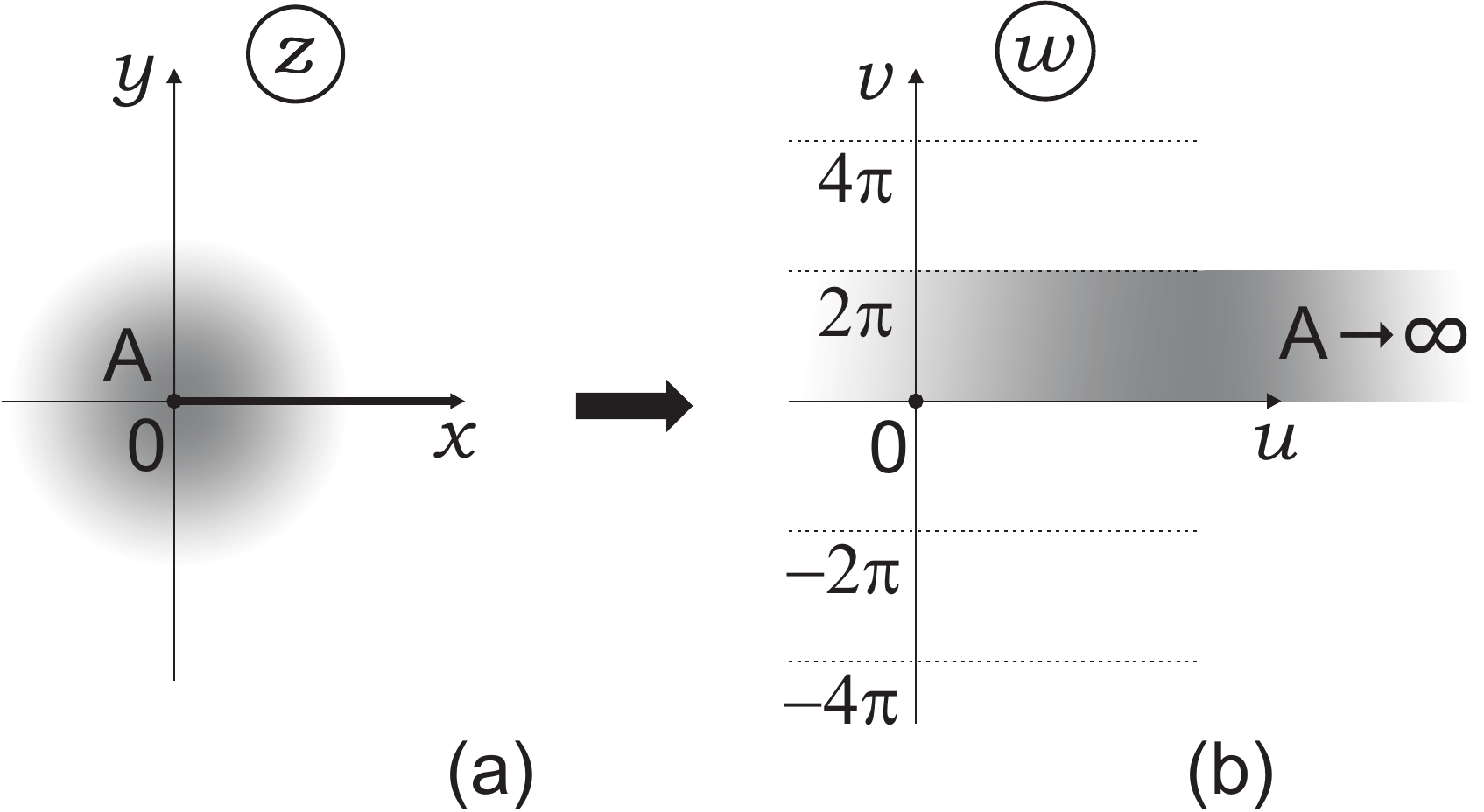}}
\caption{Conformal mapping of a plane with one branching point (obstacle) to the multi-sheet
universal covering plane.}
\label{fig:06}
\end{figure}

To demonstrate how the method of conformal mappings works, let us return to the Abelian problem and
reconsider entanglement of the random walk with the single obstacle in the plane. Place the
obstacle (the branching point) at the origin, make a cut along the positive part of the $x$-axis of
the complex plane $z=x+iy$ as shown in the \fig{fig:06} and perform a conformal transform with the
function $w(z)=\ln z$ to the universal covering space $w=u+iv$.

By elementary computations we get:
\be
\begin{cases}
u = \ln |z|\equiv \ln \rho, \quad v = \arg z; \medskip \\ |z'(w)|^{-2} = e^{-2u}
\end{cases}
\label{eq:log}
\ee
If the path ${\cal C}$ makes $n$ full turns around the branching point in $z$, it means that in the
$w$-plane the distance between extremities of the image of ${\cal C}$ along the $v$-axis is $2\pi
n$. So, closed paths crossing the cut in $z$ are transformed into the open paths in $w$. The Cauchy
problem in the lifted space is periodic in $v$ and in each Riemann sheet (the strip of width $2\pi$
in the plane $w$) can be written as follows
\be
\begin{cases}
\partial_t P(u,v,t) - D\, e^{-2u}(\partial^2_{uu} + \partial^2_{vv})P(u,v,t) \medskip \\
P(u_0,v_0,t) = \delta(u-u_0)\delta(v-v_0)\delta(t)
\end{cases}
\label{eq:cauchy}
\ee
where $\delta(...)$ is the Dirac $\delta$-function. Making use of the substitution $\rho = e^u$ and
taking into account the periodicity in $v$, we can rewrite \eq{eq:cauchy} in the following form
\be
\partial_t P(\rho,v,t) + D\left(\partial^2_{\rho\rho}+\frac{\partial_{\rho}}{\rho}+
\frac{\partial^2_{vv}}{\rho^2}\right)P(\rho,v,t) = \frac{\delta(\rho-\rho_0)}{(\rho\rho_0)^{1/2}}
\delta(v-v_0)\delta(t)
\label{eq:edw}
\ee
Seeking the solution of \eq{eq:edw} in the form
\be
P(\rho,v,t) = \sum_{m=-\infty}^{\infty} P_m e^{i m(v-v_0)}
\label{eq:edw2}
\ee
we get for $P_m$ the expression
\be
P_m = \frac{1}{\pi t a^2} e^{-\frac{\rho_0^2-\rho^2}{t a^2}}I_{|\nu+m|} \left(\frac{2\rho\rho_0}{t
a^2}\right)
\label{eq:edw3}
\ee
were we have taken into account the expression for the diffusion coefficient in form
$D=\frac{a^2}{4}$. Since
\be
\sum_{m=-\infty}^{\infty} e^{i m(v-v_0)} = 2\pi \sum_{n=-\infty}^{\infty} \delta(v+2\pi n-v_0)
\label{eq:edw4}
\ee
we arrive at \eq{eq:edw0}, where we should made the replacements $\theta \leftrightarrow v$ and
$N\leftrightarrow t$.

\begin{figure}[htbp]
\centerline{\includegraphics[width=10cm]{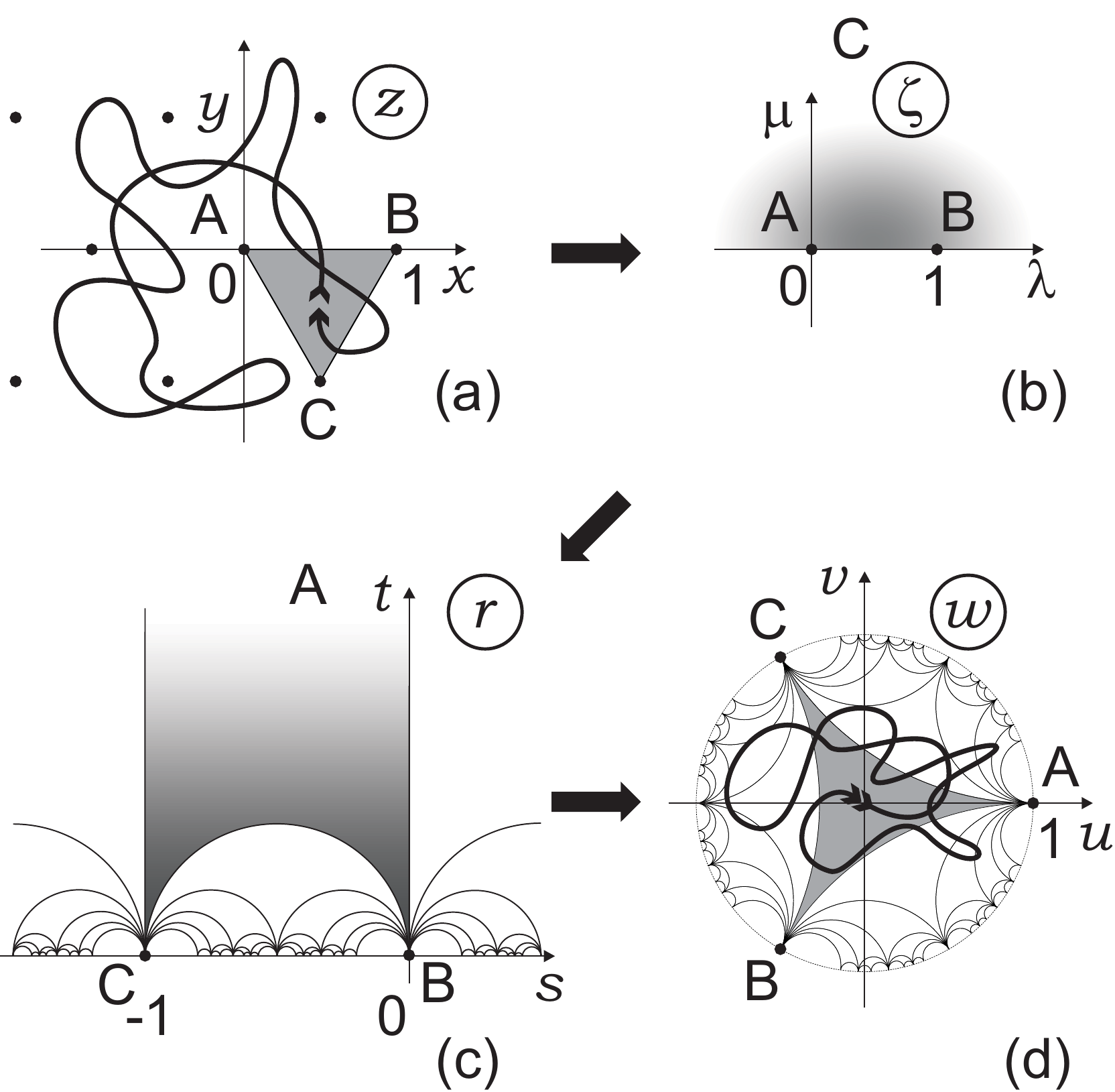}}
\caption{Conformal mapping $z(w)$ is realized as a composition of three mappings: $z(\zeta)$
[(a)--(b)], $\zeta(r)$ [(b)--(c)], and $r(w)$ [(c)--(d)]. Finally we have $z(\zeta(r(w)))$.}
\label{fig:07}
\end{figure}

Now we are in position to attack our favorite problem -- finding the probability that the random
walk of length $t$ not entangled with respect to the triangular lattice of obstacles in the complex
plane $z$, as shown in the \fig{fig:07}a. To solve the problem, we should construct the conformal
mapping of the multi-punctured plane $z$ to the universal covering space free of obstacles $w$ and
take the corresponding Jacobian of transformation, $|z(w)|^{-2}$. In this particular case finding
such a mapping is more elaborated task than for one-punctured plane, though still doable. The
conformal mapping $z(w)$ of the flat equilateral triangle $ABC$ located in $z$ onto the zero-angled
triangle $ABC$ in $w$, is constructed in three sequential steps, shown in the \fig{fig:07}a-d.

First, we map the triangle $ABC$ in $z$ onto the upper half-plane $\zeta$ of auxiliary complex
plane $\zeta$ with three branching points at 0, 1 and $\infty$ -- see the \fig{fig:07}a-b. This
mapping is realized by the function $z(\zeta)$:
\be
z(\zeta)=\frac{\Gamma(\frac{2}{3})}{\Gamma^2(\frac{1}{3})} \int_0^{\zeta}
\frac{d\xi}{\xi^{2/3}(1-\xi)^{2/3}}
\label{eq:z(zeta)}
\ee
with the following coincidence of branching points:
\be
\left\{
\begin{array}{lcl}
A(z=0)& \leftrightarrow &  A(\zeta=0)\medskip \\
B(z=1) & \leftrightarrow & B(\zeta=1) \medskip \\
C(z=e^{-i\frac{\pi}{3}}) & \leftrightarrow & C(\zeta=\infty)
\end{array} \right.
\label{eq:points1}
\ee

Second step consists in mapping the auxiliary upper half-plane $\im\zeta>0$ onto the circular
triangle $ABC$ with angles $\{\alpha,\alpha,0\}$ -- the fundamental domain of the Hecke group
\cite{hecke} in $r$, where we are intersted in the specific case $\{\alpha,\alpha,0\}=\{0,0,0\}$ --
see \fig{fig:07}b-c. This mapping is realized by the function $\zeta(r)$, constructed as follows
\cite{koppenfels}. Let $\zeta(r)$ be the inverse function of $r(\zeta)$ written as a quotient
\be
r(\zeta) = \frac{\phi_1(\zeta)}{\phi_2(\zeta)}
\label{eq:quot}
\ee
where $\phi_{1,2}(\zeta)$ are the fundamental solutions of the 2nd order differential equation of
Picard-Fuchs type:
\be
\zeta(\zeta-1) \phi''(\zeta)+\big((a+b+1)\zeta-c\big) \phi'(\zeta) + ab\phi(\zeta)=0
\label{eq:fundam}
\ee

Following \cite{caratheodory,koppenfels}, the function $r(\zeta)$ conformally maps the generic
circular triangle with angles $\{\alpha_0=\pi|c-1|,\alpha_1=\pi|a+b-c|, \alpha_\infty=\pi|a-b|\}$
in the upper halfplane of $w$ onto the upper halfplane of $\zeta$. Choosing $\alpha_{\infty}=0$ and
$\alpha_0=\alpha_1=\alpha$, we can express the parameters $(a,b,c)$ of the equation \eq{eq:fundam}
in terms of $\alpha$, taking into account that the triangle $ABC$ in the \fig{fig:07}c is
parameterized as follows $\{\alpha_0,\alpha_1,\alpha_\infty \}=\{\alpha,\alpha,0\}$ with
$a=b=\frac{\alpha}{\pi}+\frac{1}{2}, c=\frac{\alpha}{\pi}+1$. This leads us to the following
particular form of equation \eq{eq:fundam}
\be
\zeta(\zeta-1) \phi''(\zeta)+\Big(\frac{\alpha}{\pi}+1\Big)\big(2\zeta-1\big) \phi'(\zeta)
+\Big(\frac{\alpha}{\pi} +\frac{1}{2}\Big)^2\phi(\zeta)=0
\label{eq:fundam2}
\ee
where $\alpha = \frac{\pi}{m}$ and $m=3,4,...\infty$. For $\alpha=0$ Eq.\eq{eq:fundam2} takes an
especially simple form, known as Legendre hypergeometric equation \cite{golubev,hille}. The pair of
possible fundamental solutions of Legendre equation are
\be
\begin{array}{l}
\phi_1(\zeta)=F\big(\frac{1}{2},\frac{1}{2},1,\zeta\big) \medskip \\
\phi_2(\zeta)=iF\big(\frac{1}{2},\frac{1}{2},1,1-\zeta\big)
\label{eq:hyp}
\end{array}
\ee
where $F(...)$ is the hypergeometric function. From \eq{eq:quot} and \eq{eq:hyp} we get $r(\zeta) =
\frac{\phi_1(\zeta)}{\phi_2(\zeta)}$. The inverse function $\zeta(r)$ is the so-called modular
function, $k^2(r)$ (see \cite{golubev,hille,jacobi2} for details). Thus,
\be
\zeta(r) \equiv k^2(r) = \frac{\theta_2^4(0,e^{i\pi r})}{\theta_3^4(0,e^{i\pi r})}
\label{eq:zeta(r)}
\ee
where $\theta_2$ and $\theta_3$ are the elliptic Jacobi $\theta$-functions \cite{jacobi2,mum},
\be
\begin{array}{l}
\disp \theta_2\left(\chi,e^{i\pi w}\right)=2e^{i{\pi \over 4} r}
\sum_{n=0}^{\infty} e^{i\pi r n(n+1)}\cos (2n+1)\chi \medskip \\
\disp \theta_3\left(\chi,e^{i\pi r}\right)=1+2\sum_{n=0}^{\infty} e^{i\pi r n^2}\cos 2n\chi
\end{array}
\ee
and the correspondence of branching points in the mapping $\zeta(r)$ is as follows
\be
\left\{
\begin{array}{lcl}
A(\zeta=0) & \leftrightarrow & A(r=\infty) \medskip \\
B(\zeta=1) & \leftrightarrow & B(r=0) \medskip \\
C(\zeta=\infty) & \leftrightarrow & C(r=-1)
\end{array} \right.
\label{eq:points2}
\ee

Third step, realized via the function $r(w)$, consists in mapping the zero-angled triangle $ABC$ in
$r$ into the symmetric triangle $ABC$ located in the unit disc $w$ -- see \fig{fig:07}c-d. The
explicit form of the function $r(w)$ is
\be
r(w) = e^{-i\pi/3}\frac{e^{2i\pi/3}-w}{1-w}-1
\label{eq:r(w)}
\ee
with the following correspondence between branching points:
\be
\left\{
\begin{array}{lcl}
A(r=\infty) & \leftrightarrow & A(w=1) \medskip \\
B(r=0) & \leftrightarrow & B(w=e^{-2\pi i/3}) \medskip \\
C(r=-1) & \leftrightarrow & C(w=e^{2\pi i/3})
\end{array} \right.
\label{eq:points3}
\ee

Collecting \eq{eq:z(zeta)}, \eq{eq:zeta(r)}, and \eq{eq:r(w)} we arrive at the following expression
for the derivative of composite function,
\be
z'(\zeta(r(w))) = z'(\zeta)\, \zeta'(r)\, r'(w)
\ee
where $'$ stands for the derivative. We have explicitly:
$$
z'(\zeta)=\frac{\Gamma(\frac{2}{3})}{\Gamma^2(\frac{1}{3})}\,
\frac{\theta_3^{16/3}(0,\zeta)}{\theta_2^{8/3}(0,\zeta)\;\theta_0^{8/3}(0,\zeta)}
$$
and
$$
\zeta'(r)|=i\pi\frac{\theta_2^4\;\theta_0^4}{\theta_3^4}; \qquad i\frac{\pi}{4}\theta_0^4=
\frac{d}{d\zeta}\ln\left(\frac{\theta_2}{\theta_3}\right)
$$
The identity
\be
\disp \theta'_1(0,e^{i\pi\zeta})\equiv
\left.\frac{d\theta_1(\chi,e^{i\pi\zeta})}{d\chi}\right|_{\chi=0} \\ =
\pi\theta_0(\chi,e^{i\pi\zeta})\,\theta_2(\chi,e^{i\pi\zeta})\, \theta_3(\chi,e^{i\pi\zeta})
\ee
enables us to write
\be
\left|z'(r)\right|^2=h^2 \left|\theta'_1\left(0, e^{i\pi r}\right)\right|^{8/3}
\label{3:jacobian}
\ee
where $h= \left(\frac{16}{\pi}\right)^{1/3}\frac{\Gamma(\frac{2}{3})}{\Gamma^2(\frac{1}{3})}\approx
0.325$, and
\be
\theta_1(\chi,e^{i\pi r})=2e^{i\frac{\pi}{4} r} \sum_{n=0}^{\infty}(-1)^n e^{i\pi n(n+1) r}\sin
(2n+1)\chi
\label{3:ell1}
\ee
Differentiating \eq{eq:r(w)}, we get
$$
r'(w) = \frac{i\sqrt{3}}{(1-w)^2}
$$
and using this expression, we obtain the final form of the Jacobian of the composite conformal
transformation $J(z(\zeta(r(w))))$:
\be
J(z(w)) = |z'(w)|^2 = 3h^2\frac{|\eta(r(w))|^8}{|1-w|^4}
\label{eq:composite}
\ee
where
$$
\eta(r) = \big(\theta_1'(0,e^{i\pi r})\big)^{1/3}
$$
is the Dedekind $\eta$-function
\be
\eta(w)=e^{\pi i w/12}\prod_{n=0}^{\infty}(1-e^{2\pi i n w}) \quad (w=u+iv)
\label{eq:ded2}
\ee
and the function $r(w)$ is defined in \eq{eq:r(w)}.

Thus, we arrive at the diffusion equation in the unit disc $|w|<1$:
\be
\partial_t P(w,t) - \frac{a^2}{4}J(z(w)) \big(\partial^2_{uu} + \partial^2_{vv}\big) P(w,t) =
\delta(w-w_0)\delta(t)
\label{eq:modular}
\ee
with the function $J(z(w))$ given by \eq{eq:composite}. The probability to find the two-dimensional
random walk unentangled with the lattice of obstacles after time $t$, is given by the solution of
Eq.\eq{eq:modular}, where we should plug at the very end $w=w_0=0$. The probability of returning to
the same point $w_0=0$ in the initial Riemann sheet (the gray triangle in the \fig{fig:07}d)
ensures that the trajectory is: i) closed, and ii) "contractible" (i.e. topologically trivial with
respect to the lattice of obstacles).

\begin{figure}[htbp]
\centerline{\includegraphics[width=8cm]{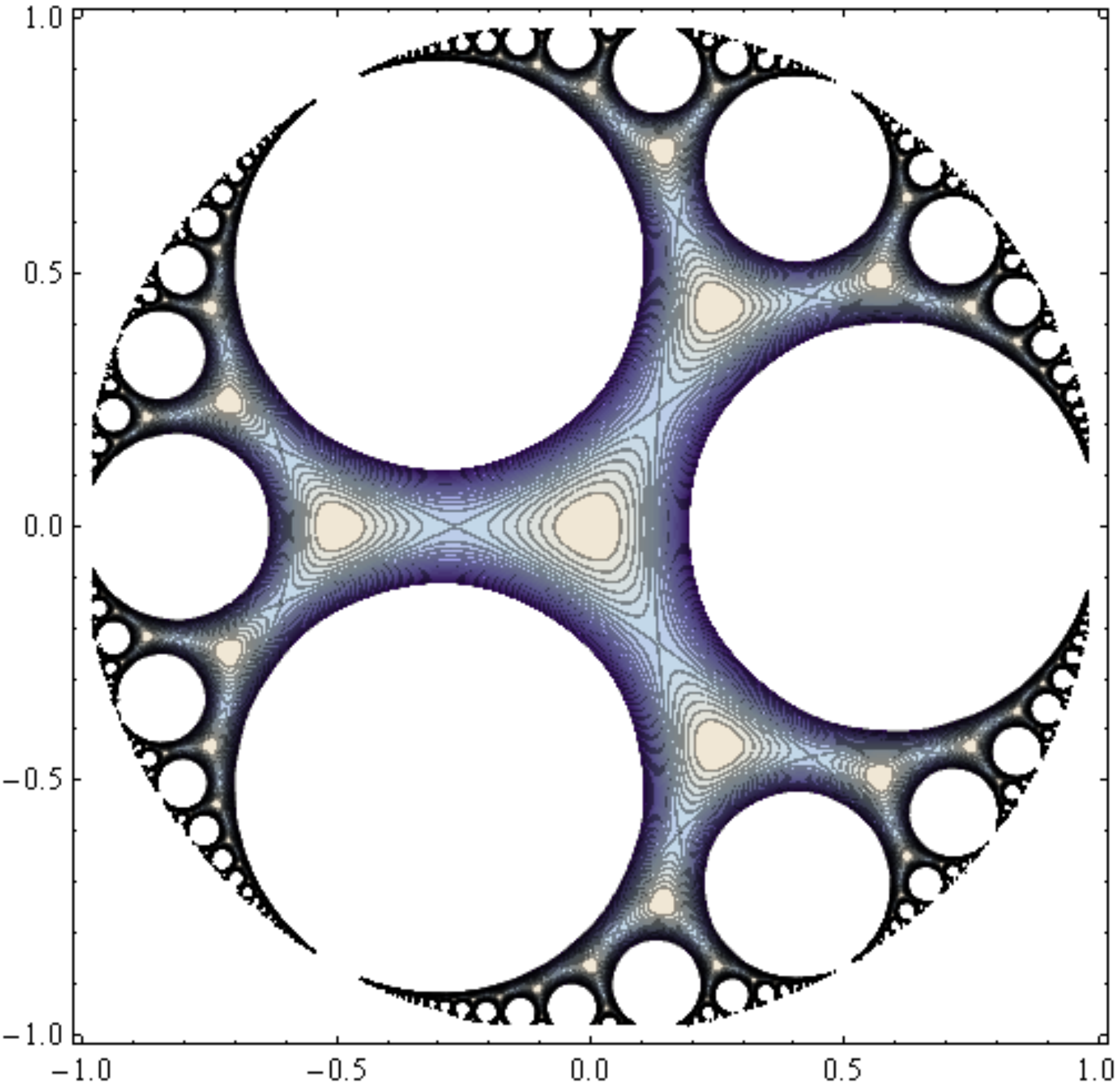}}
\caption{Density plot of the function $f(w)$ (see \eq{eq:normaliz}) above the level $f_0=0.15$.}
\label{fig:07a}
\end{figure}

Exact solution of \eq{eq:modular} is unknown, however its asymptotic behavior we can extract
relying on modular properties of Dedekind $\eta$-function \eq{eq:ded2}. Consider the normalized
Jacobian, defined as follows:
\be
f(w) = J(z(w)) \big(1-(u^2+v^2)\big)^2
\label{eq:normaliz}
\ee
where $|z|=\sqrt{u^2+v^2}$ and $\psi = {\rm arg}\, z$ are radial and angular coordinates in the
Poincare unit disc disc $|w|<1$. In the \fig{fig:07a} we have shown the density plot of the
function $f(w)$ within the unit disc $|w|<1$ for $f(w)>f_0=0.15$. As one can see from the
\fig{fig:07a}, the function $f(w)$ has identical local maxima in all the centers of circular
triangles shown in the \fig{fig:07}d. Thus, we conclude that the Jacobian $J(z(w))$ in the centers
of circular cells (domains) coincides with the metric of the Lobachevsky plane in the Poincar\'e
disc,
\be
ds^2 = \frac{du^2+dv^2}{\big(1-(u^2+v^2)\big)^2}
\label{eq:lob}
\ee
Note that the profile shown in the \fig{fig:07a} has striking similarities with the 3-branching
Cayley tree and can be interpreted as a "continuous Cayley tree".

If we slightly modify our random walk model, we can exploit the connection with the Lobachevsky
geometry. Namely, consider the random walk which stays in the vicinity of the center of a circular
cell and then rapidly jumps to the center of the neighboring cell, stays there, then jumps again,
and so forth... Such a random jumping process we can approximately describe by the diffusion in the
Poincare disc with the Lobachevsky plane metric \eq{eq:lob} and the diffusion coefficient ${\cal
D}$:
\be
\partial_t P(w,t) - {\cal D}\big((1-(u^2+v^2)\big)^2 \big(\partial^2_{uu} +
\partial^2_{vv}\big) P(w,t) = \delta(u)\delta(v)\delta(t)
\label{eq:lob2}
\ee
Making use of the change of variables $(r,\psi)\to (\rho,\psi)$, where $\rho=\ln\frac{1+r}{1-r}$,
we get the unrestricted random walk on the surface of the one-sheeted hyperboloid, ${\cal H}$
obtained by the stereographic projection from the Poincar\'e unit disc. Correspondingly the
probability $P(\rho,t)$ reads
\be
P(\rho, t)=\frac{e^{-\frac{t{\cal D}}{4}}}{4\pi\sqrt{2\pi(t{\cal D})^3}}
\int_{\rho}^{\infty}\frac{\xi \exp\left(-\frac{\xi^2}{4t{\cal D}} \right)}{\sqrt{\cosh\xi-
\cosh\rho}}d\xi
\ee

The physical meaning of the geodesic length, $\rho$, on ${\cal H}$ is straightforward: $\rho$ is
the length of the primitive path in the lattice of obstacles, i.e. the length of the shortest
trajectory remaining after all topologically allowed contractions of the random path in the lattice
of obstacles. Hence, $\rho$ can be considered a non-Abelian topological invariant, more powerful
than the Gauss linking number. This invariant is not complete except one point, $\rho=0$, where it
precisely classifies the paths belonging to the trivial homotopic class in the lattice of
obstacles.

\subsubsection{Group-theoretic methods in statistics of entangled random walks}

\begin{figure}[htbp]
\centerline{\includegraphics[width=10cm]{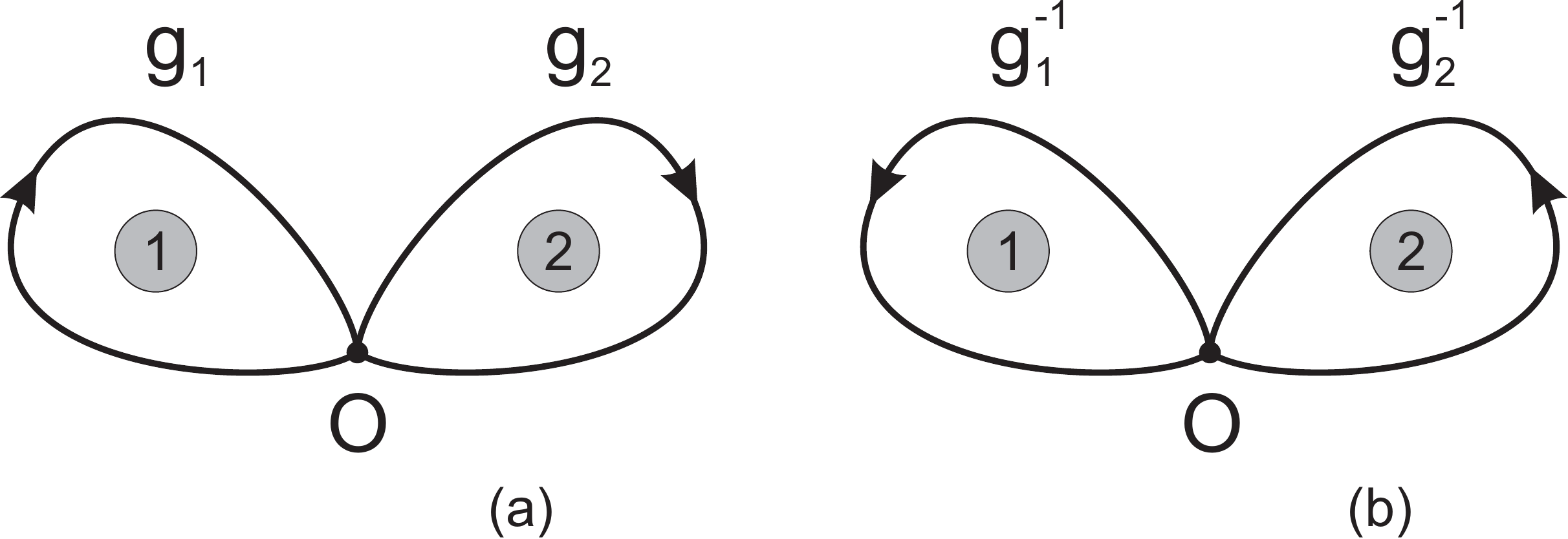}}
\caption{Generators of free group $\Gamma_2$.}
\label{fig:08}
\end{figure}

Non-Abelian entanglement of the path with two obstacles on the plane can be treated in the
group-theoretic setting. Let us associate the generators $g_1$ and $g_2$ with the clockwise full
turns around obstacles 1 and 2 respectively, and the inverse generators $g_1^{-1}$ and $g_2^{-2}$
-- around the counterclockwise turns around 1 and 2, as shown in the \fig{fig:08}. Suppose that
$g_1, g_2, g_1^{-1}, g_2^{-1}$ are the generators of the free group $\Gamma_2$ which by definition
has no commutation relations. Thus, the possible contractions in the group $\Gamma_2$ are $g_1
g_1^{-1} = g_1^{-1} g_1 = g_2 g_2^{-1} = g_2^{-1} g_2 = I$, where $I$ is the identity element.

Any closed path entangled with the obstacles 1 and 2 can be topologically presented by a word
written in terms of generators $g_1, g_2, g_1^{-1}, g_2^{-1}$. For example, the Pochhammer contour
shown in the \fig{fig:03} reads: $g_1 g_2 g_1^{-1} g_2^{-1}$. Since the group $\Gamma_2$ is
\emph{noncommutative} (non-Abelian), we have $g_1 g_2 \neq g_2 g_1$, thus we cannot exchange the
sequence of letters and replace $g_1 g_2$ by $g_2 g_1$. However, in the \emph{commutative}
(Abelian) group generated by the set $\{f_1, f_2, f_1^{-1}, f_2^{-1}\}$, we can do so, since $f_1
f_2 = f_2 f_1$, and the Pochhammer contour in the Abelian representation becomes contractible: $f_1
f_2 f_1^{-1} f_2^{-1} = f_2 \underbracket{f_1 f_1^{-1}}_I f_2^{-1} = \underbracket{f_2 f_2^{-1}}_I
= I$.

\begin{figure}[htbp]
\centerline{\includegraphics[width=12cm]{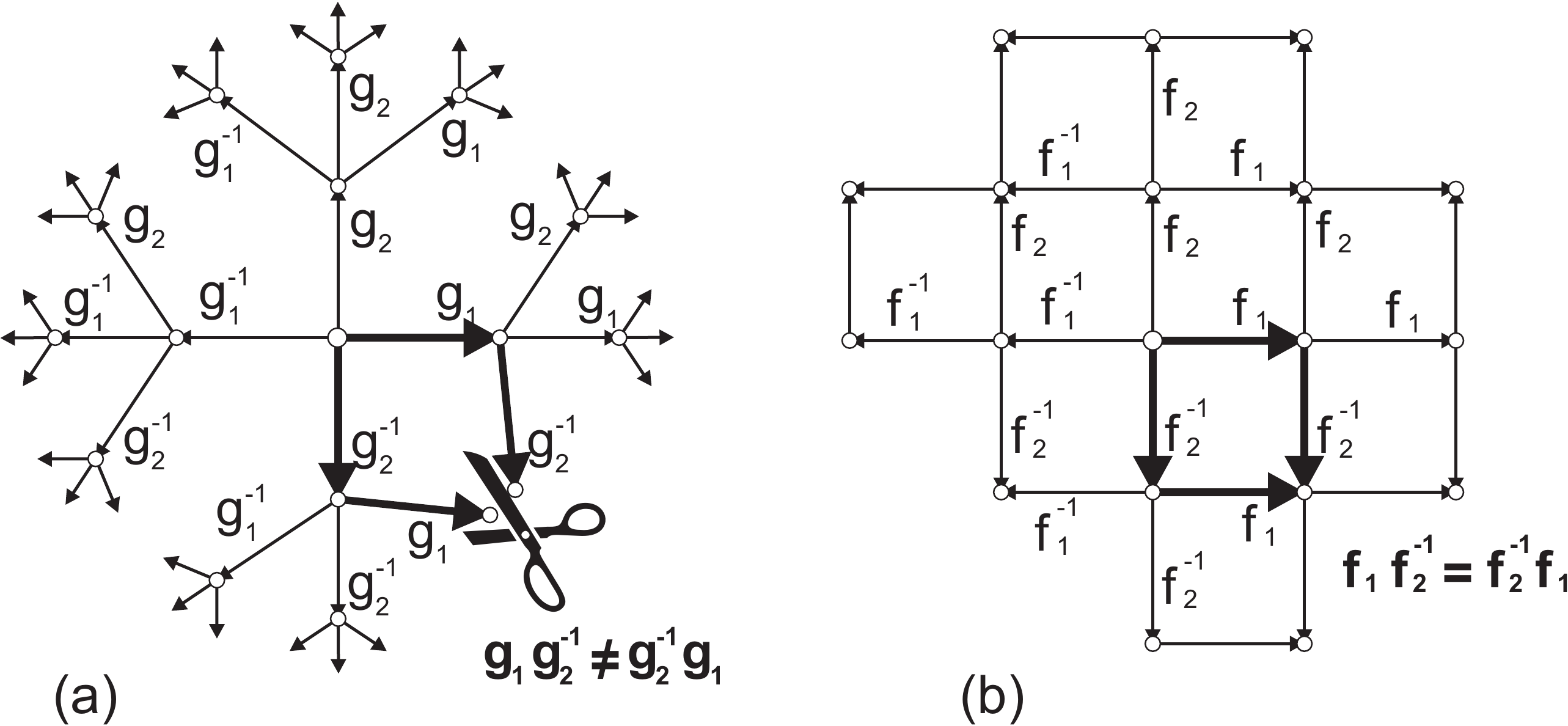}}
\caption{Visualization of commutation relations in commutative (a) and in free (b) groups.}
\label{fig:09}
\end{figure}

Suppose now that we have a set ${\cal S}=\{g_1, g_2, g_1^{-1}, g_2^{-1}\}$ and rise random words of
$N$ letters by sequential adding of generators from the set ${\cal S}$. Each generator in ${\cal
S}$ we take with the probability $p=\frac{1}{4}$. We are interested in computing the partition
function $Z_N(x)$ for all $N$-letter random words to have the irreducible ("primitive") word of $x$
letters. The partition function $Z_N(0)$ gives the number of $N$-letter words that are completely
reducible (i.e. unentangled) with the obstacles 1 and 2. The word counting problem in the free
group $\Gamma_2$ can be visualized as the trajectories (built by sequential adding of letters) on
the 4-branching Cayley tree shown in the \fig{fig:09}a. The irreducible (primitive) word is the
shortest "bare" path along the tree connecting the extremities of the trajectory.

The partition function, $Z_N(x)$, of all $N$-step paths on the 4-branching Cayley tree, starting at
the origin and ending at some distance $x$ from it, satisfies the following recursion relation
\be
\left\{\begin{array}{rcll} \disp Z_{N+1}(x) & = & 3Z_N(x-1) + Z_N(x+1), & \quad x\ge 2
\medskip \\ \disp Z_{N+1}(x) & = & 4 Z_N(x-1) + Z_N(x+1), & \quad x=1 \medskip \\
Z_{N+1}(x) & = & Z_N(x+1), & \quad x=0 \medskip \\
Z_{N}(x) & = & 0, & \quad x\le -1 \medskip \\
Z_{N=0}(x) & = & \delta_{x,0},
\end{array} \right.
\label{eq:t-04}
\ee
where $x$ is the distance (the bare path) from the root of the Cayley graph, measured in number of
generations of the tree. By making a shift $x\to x+1$, one can rewrite \eq{eq:t-04} as
\be
\left\{\begin{array}{rcll} Z_{N+1}(x) & = & 3 Z_N(x-1)+Z_N(x+1) + \delta_{x,2}\, Z_{N}(x-1), &
\qquad x\ge 1
\medskip \\
Z_{N}(x) & = & 0,  & \qquad x=0 \medskip \\
Z_{N=0}(x) & = & \delta_{x,1},
\end{array} \right.
\label{eq:t-05}
\ee
where $\delta_{x,y}$ is the Kronecker $\delta$--function: $\delta_{x,y}=1$ for $x=y$ and
$\delta_{x,y}=0$ for $x\neq y$.

Equation \eq{eq:t-05} can be symmetrized by the substitution
\be
Z_N(x) = A^N B^x W_N(x).
\label{eq:t-zw}
\ee
Selecting $A=B=\sqrt{3}$, we arrive at the equation
\be
\left\{\begin{array}{rcll} W_{N+1}(x) & = & \disp W_N(x-1) + W_N(x+1) +
\frac{1}{3}\,\delta_{x,2}\,W_N(x-1), & \qquad x\ge 1 \medskip \\
W_n(x) & = & 0, & \qquad x=0 \medskip \\
W_{N=0}(x) & = & \disp \frac{\delta_{x,1}}{\sqrt{3}}
\end{array} \right.
\label{eq:t-06}
\ee

Introducing the generating function
\be
{\cal  W}(s,x) = \sum_{N=0}^{\infty}W_N(x) s^n \qquad \left(W_N(x) = \frac{1}{2\pi i}\oint {\cal
W}(s,x) s^{-N-1}\, ds \right)
\label{eq:t-07}
\ee
and its $\sin$--Fourier transform
\be
\tilde{{\cal  W}}(s,q) = \sum_{x=0}^{\infty}{\cal  W}(s,x) \sin qx \qquad \left({\cal  W}(s,x) =
\frac{2}{\pi} \int_{0}^{\pi} \tilde{{\cal  W}}(s,q) \sin qx\, dq \right),
\label{eq:t-08}
\ee
one obtains from \eq{eq:t-06}
\be
\frac{\tilde{{\cal  W}}(s,q)}{s} - \frac{\sin q}{s\sqrt{p-1}} = 2\cos q\, \tilde{{\cal  W}}(s,q) +
\frac{2}{\pi} \frac{1}{3} \sin 2q \int_{0}^{\pi} \tilde{{\cal W}}(s,q) \sin q\, dq.
\label{eq:t-09}
\ee
Solving \eq{eq:t-09} and performing the inverse Fourier transform, we arrive at the following
explicit expression for the generating function ${\cal W}(s,x)$:
\begin{multline}
{\cal W}(s,x)=\frac{2}{\pi}\int_0^{\infty}\tilde{{\cal W}}(s,q) \sin qx\, dq \\ =
\frac{1}{s\sqrt{3}}\left(\frac{1-\sqrt{1-4s^2}}{2s}\right)^x \left(1+
\frac{2\left(1-\sqrt{1-4s^2}\right)}{12s^2-1\left(1-\sqrt{1-4s^2}\right)^2}\right).
\label{eq:t-14}
\end{multline}
Since, by definition, $Z_N(x) = A^N B^x W_N(x)$ (see \eq{eq:t-zw}), we can write down the relation
between the generating functions of $Z_N(x)$ and of $W_N(x)$:
\be
{\cal Z}(\lambda,x)=\sum_{N=0}^{\infty} Z_N(x) \lambda^N = \sum_{N=0}^{\infty} A^N B^x W_N(x)
\lambda^N=B^x {\cal W}(\lambda A,x).
\ee
Thus,
\be
{\cal Z}(\lambda,x) = 3^{x/2}{\cal W}(\lambda \sqrt{3},x),
\label{eq:t-15}
\ee
where ${\cal W}(\lambda \sqrt{3},x)$ is given by \eq{eq:t-14} and we should substitute $\lambda
\sqrt{3}$ for $s$. The partition function, ${\cal Z}(\lambda,x)$, of the random walk ensemble reads
\be
{\cal Z}(\lambda,x) = \frac{6 \lambda \left(\frac{\disp 1 - \sqrt{1 - 4 \lambda^2(p-1)}}{\disp
2\lambda} \right)^x}{18\lambda^2-\left(1-\sqrt{1-12\lambda^2}\right)}.
\label{eq:t-16}
\ee
For the grand partition function of all trajectories returning to the origin, ${\cal
Z}(\lambda)\equiv {\cal Z}(\lambda,x=0)$, we get the following expression
\be
{\cal Z}(\lambda) = \frac{6 \lambda}{18\lambda^2-\left(1-\sqrt{1-12\lambda^2}\right)}.
\label{eq:t-17}
\ee
To extract the asymptotic behavior of the partition function $Z_N$ (the number of trajectories
returning to the origin after $N$ steps, one should perform the inverse transform similar to
\eq{eq:t-07}
\be
Z_N = \frac{1}{2\pi i}\oint {\cal Z}(\lambda)\lambda^{-N-1}d \lambda \sim
\frac{\left(2\sqrt{3}\right)^N}{N^{3/2}}
\label{eq:t-17a}
\ee
As it should be, the probability to return to the origin on a 4-branching Cayley tree, $Z_N/(4^N)$
is exponentially small.

\begin{figure}[htbp]
\centerline{\includegraphics[width=12cm]{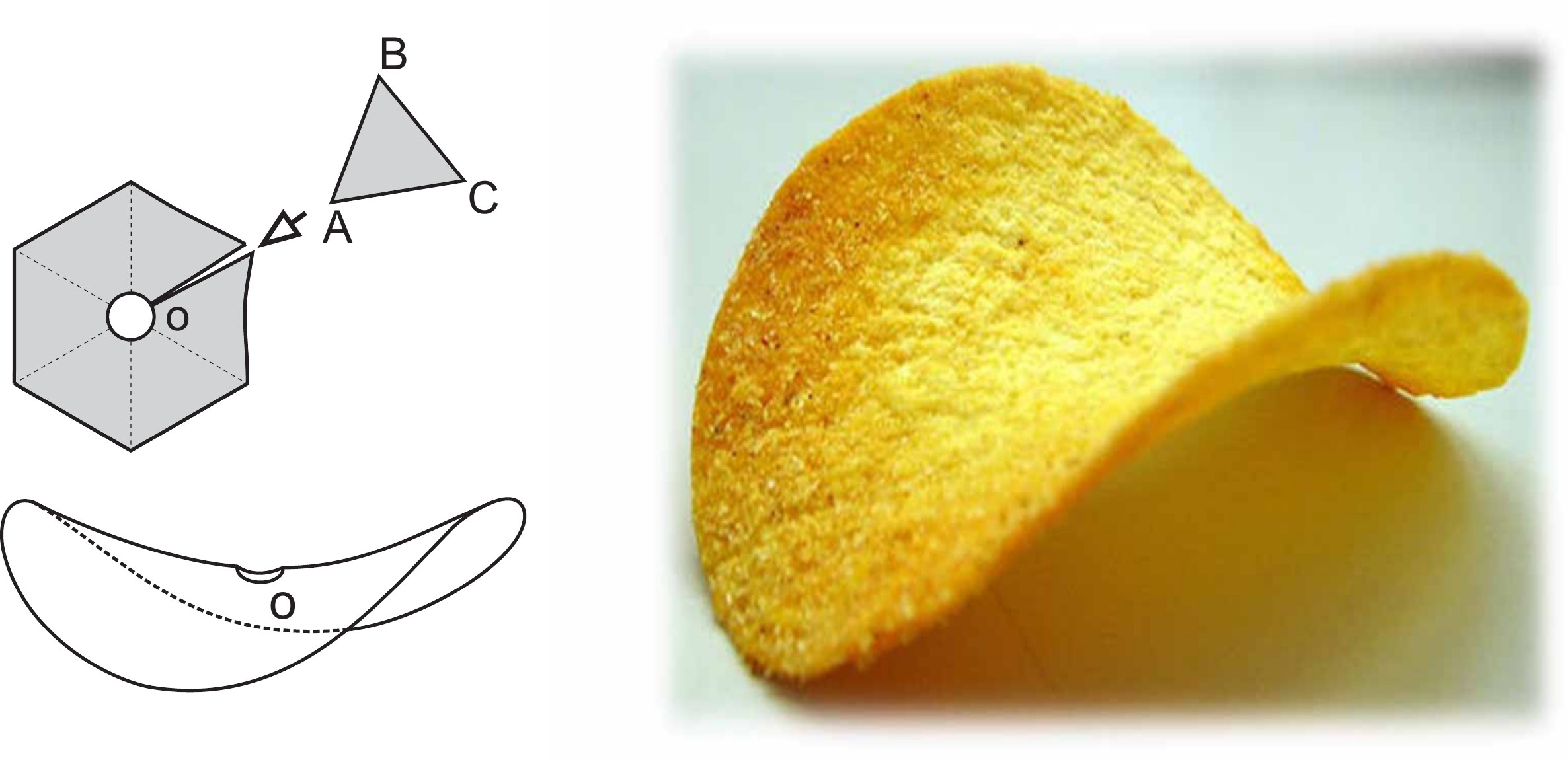}}
\caption{Isometric embedding of the free group into the Lobachevsky plane (Riemann surface
of constant negative curvature).}
\label{fig:10}
\end{figure}

Let us note some striking topological similarity between the Cayley tree structure of the
noncommutative group $\Gamma_2$ shown in the \fig{fig:08} and the metric structure of the modular
group, visualized in the \fig{fig:07}. This similarity is not occasional. Having the graph of the
group $\Gamma_2$, we can ask the question in which Riemann surface the graph of the group
$\Gamma_2$ can be isometrically embedded. The answer is that the Cayley tree is the graph of
isometries (one of many) of the Lobachevsky plane (the Riemann surface of constant negative
curvature). This is schematically depicted in the \fig{fig:10} where the chips (the saddle) is the
example of the surface with constant negative Gaussian curvature. Contrary to that, the commutative
group $\{f_1, f_2, f_1^{-1}, f_2^{-1}\}/[f_1 f_2 = f_2 f_1]$ isometrically covers the planar square
lattice -- see the \fig{fig:09}b.

\subsubsection{Conditional Brownian bridges in Hyperbolic spaces}

The result formulated in this Section is the central point, connecting statistics of random
walks in Hyperbolic spaces and the topology of knotted random walks. There are few different
incarnations of one and the same question concerning the conditional return probability of the
symmetric random walk in the Hyperbolic geometry:

(i) For the problem of the  conditional paths counting on the Cayley tree, we are interested in the
following question. Let $Z_N(x)$ be the number of $N$-step path on the Cayley tree starting from
the origin and ending at some distance $x$ measured in number of the tree generations
("coordinational spheres") from the root point. Let the paths starting from the tree root, reach
the distance $x$ after $M$ steps, and then return to the origin at the very last step, $N$. The
corresponding conditional distribution, $P(x,M,N)$ we can compute as follows
\be
P(x,M,N) = \frac{Z_M(x) Z_{N-M}(x)}{Z_M(0) V(x)}
\label{eq:c-01}
\ee
The expression \eq{eq:c-01} means that the entire $N$-step Brownian bridge (the path returning to
the origin) consists of two independent parts: the $M$-step part of the path form the root point to
\emph{some} point located at the distance $x$, and $N-M$-step part of the path from the root point
again to \emph{some} point located at the distance $x$. Now we have to ensure that the ends of
these two $M$-- and $N-M$--parts \emph{coincide} at the point ${\bf x}$. The factor $V(x)$ in the
denominator of \eq{eq:c-01} is the number of different points located at the distance $x$ from the
root of the tree, $V(x) = 4\times 3^{x-1}$ for the 4-branching Cayley tree. Thus, $Z_M(x)
V^{-1}(x)$ is the probability that the $M$-step part ends in some \emph{specific} point ${\bf x}$.
Substituting expressions for $Z_M(x)$, $Z_{N-M}(x)$, $Z_N(0)$ in \eq{eq:c-01}, we arrive at the
following asymptotic form of $P(x,M,N)$ at $x\gg 1$:
\be
P(x,M,N) \sim \sqrt{\frac{N}{2\pi M(N-M)}} e^{x^2\left(\frac{1}{2M}+\frac{1}{2(N-M)}\right)}
\label{eq:c-02}
\ee
Computing $\left<x^2\right>$ with the function $P(x,M,N)$, we get
\be
\bar{x}=\sqrt{\left<x^2\right>} = \frac{\sum_{x=0}^{\infty}x^2 P(x,M,N)}{\sum_{x=0}^{\infty}}
= \sqrt{\frac{M(N-M)}{N}}
\label{eq:c-03}
\ee
As one sees, for any $M=cN$ ($c={\rm const}$, $0<c<1$), the average distance $\bar{x}$ has the
behavior typical for an ordinary random walk,
\be
\bar{x}=\sqrt{a(1-a)}\sqrt{N}
\label{eq:c-04}
\ee
Thus, the typical behavior of intermediate points of the Brownian bridge on the Cayley tree is
statistically the same as on the one-dimensional lattice, i.e. the drift on the tree which occurs
due to asymmetry of the random walk (the probability to go \emph{from} the root is larger than the
probability to come \emph{back} to the root) is completely compensated. This is the key point for
existence of topologically nontrivial crumpled globule structure, which will be discussed in the
following Section.

(ii) Consider now the asymptotics of the conditional Brownian bridge in the Lobachevsky plane.
Construct the desired conditional probability, $W(x,M,N)$ as follows
\be
W(\rho,\tau,t) = \frac{P(\rho, \tau) P(\rho, t-\tau)}{P(0,t)} v(\rho)
\label{eq:c-05}
\ee
where $P(\rho,t)$ defines the probability that the random path after time $t$ ends in the
\emph{specific} point located at the distance $\rho$ of the Labachevsky plane, and
$v(\rho)=\sinh\rho$ is the circumference of circle of radius $\rho$ in the Lobachevsky plane.

The diffusion equation for the density $P({\bf q},t)$ of the free random walk on a Riemann manifold
is governed by the Beltrami-Laplace operator:
\be
\partial_t P({\bf q},t) = {\cal D} \frac{1}{\sqrt{g}}\frac{\partial}{\partial q_i}
\left(\sqrt{g} \left(g^{-1}\right)_{ik}\frac{\partial}{\partial q_k}\right) P({\bf q},t)
\label{eq:lob3}
\ee
where
\be
\begin{array}{c}
\disp P({\bf q},t=0)=\delta({\bf q}) \medskip \\
\disp \int \sqrt{g} P({\bf q},t) d{\bf q} = 1
\end{array}
\ee
and $g_{ik}$ is the metric tensor of the manifold; $g=\det g_{ik}$. For the Lobachevsky plane one
has
\be
||g_{ik}||=\left|\left| \begin{array}{cc} 1 & 0 \medskip \\ 0 & \sinh^2 \rho \end{array}
\right|\right|
\ee
where $\rho$ stands for the geodesics length in the Lobachevsky plane. The corresponding diffusion
equation now reads
\be
\partial_t P(\rho,\varphi,t)= {\cal D}\left(\partial^2_{\rho\rho}+\coth\rho \partial_{\rho} +
\frac{1}{\sinh^2\rho} \partial^2{\varphi\varphi}\right) P_p(\rho,\varphi,t)
\label{eq:lob4}
\ee
The radially symmetric solution of Eq.(\ref{eq:lob4}) is
\be
\begin{array}{lll}
P(\rho,t) & = & \disp \frac{e^{-\frac{t{\cal D}}{4}}}{4\pi\sqrt{2\pi(t{\cal D})^3}}
\int_{\rho}^{\infty}\frac{\xi \exp\left(-\frac{\xi^2}{4t{\cal
D}}\right)} {\sqrt{\cosh \xi - \cosh \rho}} d\xi \medskip \\
& \simeq & \disp \frac{e^{-\frac{t{\cal D}}{4}}}{4\pi t{\cal D}}\left(\frac{\rho}
{\sinh\rho}\right)^{1/2} \exp\left(-\frac{\rho^2}{4t{\cal D}}\right)
\end{array}
\label{eq:lob5}
\ee
(compare to \eq{eq:t-14}). Substituting \eq{eq:lob5} into \eq{eq:c-05}, we get for the conditional
probability $W(\rho,\tau,t)$ the following asymptotic expression
\be
W(\rho,\tau,t) = \frac{N}{4\pi{\cal D}\tau(t-\tau)}\rho \exp\left\{-\frac{\rho^2} {4{\cal
D}}\left(\frac{1}{\tau}+\frac{1}{t-\tau}\right)\right\}
\label{eq:lob6}
\ee
Hence we again reproduce the Gaussian distribution function with zero mean.

(iii) Equations \eq{eq:c-02} and \eq{eq:lob6} describing the conditional distributions Brownian
bridges on the Cayley tree and on the Riemann surface of constant negative curvature, have direct
application to the conditional distributions of Lyapunov exponents for products of non-commutative
matrices. Consider for specificity the random walk on the group $SL(2,R)$. Namely, we multiply
sequentially $N$ random matrices $M_j=\left(\begin{array}{cc} a_j & b_j \\ c_j & d_j
\end{array}\right) \in SL(2,R)$, whose entries $\{a_i,b_j,c_j,d_j\}$ are randomly distributed for
any $j=1,...,N$ in some finite support subject to the relation $a_j d_j-b_j c_j=1$. Thus we have a
product of random matrices
\be
Q(N) = \left(\begin{array}{cc} a_1 & b_1 \\ c_1 & d_1 \end{array}\right) \left(\begin{array}{cc}
a_2 & b_2 \\ c_2 & d_2 \end{array} \right) ... \left(\begin{array}{cc} a_M & b_M \\ c_M & d_M
\end{array}\right) ... \left(\begin{array}{cc} a_N & b_N \\ c_N & d_N
\end{array}\right)
\label{eq:c-pr1}
\ee
The asymptotic $N\to\infty$ behavior of the largest eigenvalue, $\Lambda_N$ of the typical (i.e.
averaged over different samples) value of $Q(N)$ is ensured by the F\"urstenberg theorem
\cite{furstenberg1963noncommuting}, which states that
\be
\Lambda_N\sim e^{\delta_1 N}
\label{eq:fuerst}
\ee
where $\delta_1$ is the group-- and measure--specific, though $N$-independent "Lyapunov exponent".

Motivated by examples (i) and (ii), consider the conditional Brownian bridge in the space of
matrices, i.e. consider such products $Q(N)$ that are equal to the unit matrix and ask about the
typical behavior of the largest eigenvalue of first $M$ as shown below:
\be
\overbracket{\left(\begin{array}{cc} a_1 & b_1 \\ c_1 & d_1 \end{array}\right)
\left(\begin{array}{cc} a_2 & b_2 \\ c_2 & d_2 \end{array}\right) ... \left(\begin{array}{cc} a_M &
b_M \\ c_M & d_M \end{array}\right)}^{\Lambda_M=?} ...  \left(\begin{array}{cc} a_N & b_N \\ c_N &
d_N
\end{array}\right) = \left(\begin{array}{cc} 1 & 0 \\ 0 & 1 \end{array} \right)
\label{eq:c-pr2}
\ee
The answer to this question is as follows: $\Lambda_M = e^{\delta_2 \sqrt{\frac{M(N-M)}{N}}}$ i.e.
for $M=cN$ ($0<c<1$), one has
\be
\Lambda_M = e^{\delta_3 \sqrt{N}}
\label{eq:c-pr3}
\ee
where we have absorbed the constants in $\delta_3=\delta_2 \sqrt{a(1-a)}$. The proof of the
corresponding theorem using the method of large deviations, can be found in \cite{nesin2}, however
the result \eq{eq:c-pr3}, which holds for random walks on any noncommutative group, is easy to
understand qualitatively. It is sufficient to recall that the Lobachevsky plane $H$ can be
identified with the group $SL(2,R)/SO(2)$. Thus, the Brownian bridge on the group $SL(2,R)/SO(2)$,
can be viewed either as the conditional random walk governed by the Beltrami-Laplace operator, or
as the product of random matrices.

We arrive at the following conclusion. The Brownian bridge condition for random walks in the space
of constant negative curvature makes the curved space effectively flat and turns the corresponding
conditional distribution for the intermediate time moment of random walks to the Gaussian
distribution with zero mean. This result is very general and can be applied to random walks on
various noncommutative groups, such as modular group, $SL(n,R)$, braid groups $B_n$, etc. This
result is crucial in our further discussions of the crumpled state of collapsed unknotted polymer
chain.

\subsection{Crumpled globule: Topological correlations in collapsed unknotted rings}

In 1988 we have theoretically predicted the new condensed state of a ring unentangled and unknotted
macromolecule in a poor solvent. We named this state "the crumpled globule" and studied its unusual
fractal properties \cite{gns}. That time our arguments were rather hand-waving and more solid
understanding came essentially later, around 2005 \cite{vasne,jktr}. As we shall show, the most
striking physical arguments of the estimation of the degree of entanglement of a part of a long
unknotted random trajectory confined in a small box are provided by the statistics of conditional
Brownian bridges in the space of constant negative curvature.

First of all one has to define the topological state of a part of a ring polymer chain. As we
discussed in the Introduction, mathematically rigorous definition of the topological state exists
for closed or infinite paths only. However, the everyday's experience tells us that open but
sufficiently long rope can be knotted. Hence, it is desirable to introduce a concept of a
\emph{quasiknot} available for topological description of open paths. For the first time the idea
of quasiknots in a polymer context had been formulated by I.M. Lifshits and A.Yu. Grosberg
\cite{ligr}. They argued that the topological state of a linear polymer chain in a collapsed
(globular) state is defined much better than topological state of a random coil. Actually, the
distance between the ends of the chain in a globule is of order $R \sim a N^{1/3}$, where $a$ is a
size of a monomer and $N$ is a number of monomers in a chain. Taking into account that $R$ is
sufficiently smaller than the contour length $N$ and that the density fluctuations in the globular
state are negligible, we may define the topological state of a path in a globule as a topological
state of composite trajectory consisting of a chain itself and a segment connecting its ends. This
composite structure can be regarded as a quasiknot for an open chain in a collapsed state. Later we
shall repeatedly use this definition.

\begin{figure}[htbp]
\centerline{\includegraphics[width=10cm]{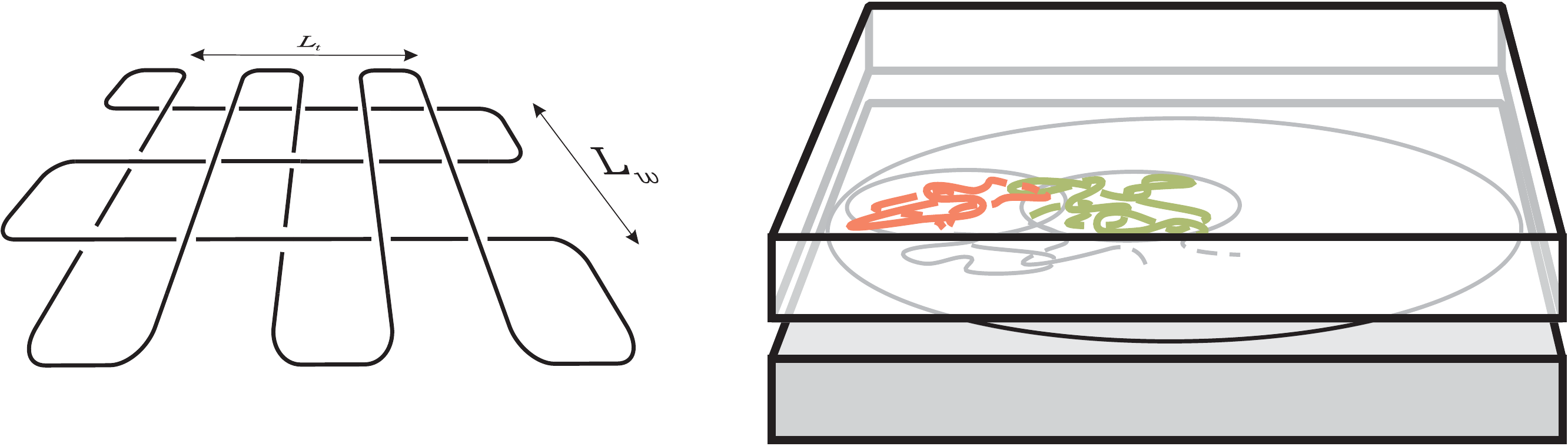}}
\caption{(a) Random woven carpet corresponding to the trivial knot; (b) Dense knot
confined in a thin slit.}
\label{fig:11}
\end{figure}

The influence of topological constraints on statistical properties of polymers, namely, the random
knotting probability, in confined geometries has been numerically considered in \cite{tesi}, while
the paper \cite{orlan} has been devoted to the determination of the equilibrium entanglement
complexity of polymer chains in melts. In the works \cite{jktr} we made further step, considering
the topological state of a part of ring unknotted polymer chain in a confined geometry, where the
compact configuration of the polymer is modelled by dense lattice knots. The knot is called "dense"
if its projection onto the plane completely fills the rectangle lattice $M$ of size $L_{v} \times
L_{h}$ as it is shown in the \fig{fig:11}a, resembling the celtic knot in the \fig{fig:01}. The
lattice $M$ is filled densely by a single thread, which crosses itself in each vertex of the
lattice in two different ways: "up" or "down". The topology of a lattice diagram is defined by the
up-down passages, and by the prescribed boundary conditions. The "woven carpet" shown in
\fig{fig:11}a corresponds to a trivial knot. To avoid any possible confusions, we apply our model
to the polymer ring located in a thin slit between two horizontal plates as it is shown in
\fig{fig:11}b. It is evident that the ring chain in a thin thin slit becomes a quasi
two-dimensional system. Our lattice model is oversimplified (even for the polymer chain in a thin
slit) because it does not take into account the spatial fluctuations of a knotted polymer chain.
However, we expect that our model properly describes the condensed (globular) structure of a
polymer ring because the chain fluctuations in the globule are essentially suppressed and the chain
has reliable thermodynamic structure with a constant density \cite{ligr}.

We are interested in the following statistical-topological question inspired by the conditional
Brownian bridge ideology. Define at each intersection of vertical and horizontal threads (i.e. in
each "lattice vertex") $k$ the random variable $\varepsilon_k$, taking values
\be
\varepsilon_k= \begin{cases} +1 & \mbox{for "up" crossing} \\ -1 & \mbox{for"down"
crossing} \end{cases}
\label{eq:crump1}
\ee
The set of independently generated quenched random variables $\{\varepsilon_1,..., \varepsilon_N\}$
in all vertices of the lattice diagram, together with the boundary conditions, define the knot
topology.

\begin{figure}[htbp]
\centerline{\includegraphics[width=10cm]{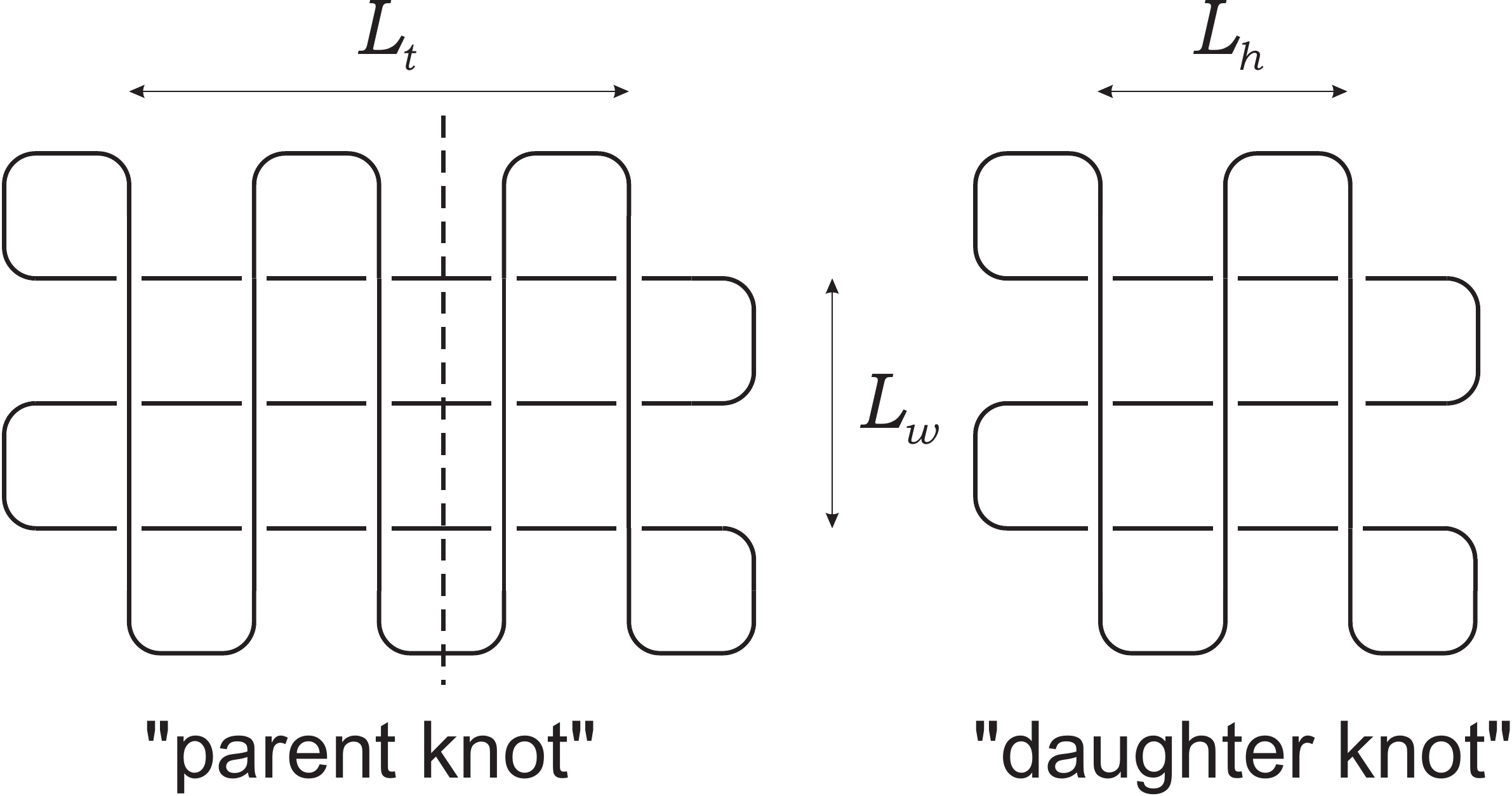}}
\caption{(a) Trivial "parent" knot; (b) "Daughter" knot obtained by cutting a part of
the parent knot.}
\label{fig:11a}
\end{figure}

Suppose that we consider such sub-ensemble of crossings $\{\varepsilon_1,..., \varepsilon_N\}$ that
corresponds to the trivial entire ("parent") knot. Let us cut a part of a parent trivial knot and
close open ends of the threads as it is shown in \fig{fig:11a}. This way we get the well defined
"daughter" quasiknot. We are interested in the typical topological state of daughter quasiknots
under the condition that the parent knot is trivial. I think, the reader can feel in this
formulation the flavor of conditional Brownian bridges...

The averaged knot complexity, $n$, understood as the minimal number of crossings on the knot
diagram, for the unconditional random knotting of knot diagram behaves as $n\sim N$ where $N$ is
the initial size of the lattice knot. By the semi-analytic and semi-numeric arguments we have shown
in \cite{jktr} that the typical conditional complexity, $n^*$, of the daughter knot of size $M$
behaves as $n^* \sim \sqrt{\frac{M(N-M)}{N}}$ and for $M=cN$ $(0<c<1)$, $N\gg 1$, has the
asymptotic behavior
\be
n^* \sim \sqrt{a(1-a)} \sqrt{N}
\label{eq:unknot}
\ee
Thus, each macroscopic part of a dense lattice trivial knot is weakly knotted (compared to the
unconditional random knotting).

In 2015 we have performed in \cite{nech-mirny} extensive Monte-Carlo simulations for self-avoiding
polymer chains in confined geometry in 3D space. The role of topological constraints in the {\em
equilibrium} state of a single compact and unknotted  polymer remains unknown. Previous studies
\cite{gns,gr-rab} have put forward a concept of the {\em crumpled globule} as the equilibrium state
of a compact and unknotted polymer. In the crumpled globule, the subchains were suggested to be
space-filling and unknotted. Recent computational studies examined the role of topological
constraints in the non-equilibrium (or quasi-equilibrium) polymer states that emerge upon polymer
collapse \cite{mirny,rost,obukhov2,chu,chertovich2014crumpled}. This non-equilibrium state, often
referred to as the \emph{fractal globule} \cite{mirny,leonid_review}, can indeed possess some
properties of the conjectured equilibrium crumpled globule. The properties of the fractal globule,
its stability \cite{schiessel}, and its connection to the equilibrium state are yet to be
understood.

Elucidating the role of topological constraints in equilibrium and non-equilibrium polymer systems
is important for understanding the organization of chromosomes. Long before experimential data on
chromosome organization became available \cite{mirny}, the crumpled globule was suggested as a
state of long DNA molecules inside a cell \cite{gr-rab}. Recent progress in microscopy
\cite{cremer2011review} and genomics \cite{dekker_NRG} provided new data on chromosome organization
that appear to share several features with topologically constrained polymer systems
\cite{rosa_everaers,mirny,grosberg2012review}. For example, segregation of chromosomes into
territories resembles segregation of space-filling rings \cite{gr-krem2,smrek,stasiak}, while
features of intra-chromosomal organization revealed by Hi-C technique are consistent with a
non-equilibrium fractal globule emerging upon polymer collapse \cite{mirny,leonid_review,machine}
or upon polymer decondensation \cite{rosa_everaers}. These findings suggest that topological
constraints can play important roles in the formation of chromosomal architecture
\cite{halverson2014melt}.

In \cite{nech-mirny} we have examined the role of topological constraints in the equilibrium state
of a compact polymer. We performed the equilibrium Monte Carlo simulations of a confined
unentangled polymer ring with and without topological constraints. Without topological constraints,
a polymer forms a classical equilibrium globule with a high degree of knotting
\cite{grosberg-khokhlov,knots-kardar, knots-kardar2}. A polymer is kept in the globular state by
impenetrable boundaries, rather than pairwise energy interactions, allowing fast equilibration at a
high volume density.

In the work \cite{nech-mirny} we found that topological states of closed subchains (loops) are
drastically different in the two types of globules and reflect the topological state of the whole
polymer. Namely, loops of the unknotted globule are only weakly knotted and mostly unconcatenated.
We also found that spatial characteristics of small knotted and unknotted globules are very
similar, with differences starting to appear only for sufficiently large globules. Subchains of
these large unknotted globules become asymptotically compact ($R_G(s)\sim s^{1/3}$), forming
crumples. Analysis of the fractal dimension of surfaces of loops suggests that crumples form
excessive contacts and slightly interpenetrate each other. Overall, in the the asymptotic limit
(for very long chains) we have supported the conjectured crumpled globule concept \cite{gns}.
However, the results of \cite{nech-mirny} also demonstrate that the internal organization of the
unknotted globule at equilibrium differs from an idealized hierarchy of self-similar isolated
compact crumples.

In the simulations a single homopolymer ring with excluded volume interactions was modelled on a
cubic lattice and confined into a cubic container at a volume density $0.5$. The Monte Carlo method
with non-local moves (see \cite{nech-mirny} and references therein) allowed us to study chains up
to $N=256\,000$. If monomers were prohibited to occupy the same site, this Monte Carlo move set
naturally constrains topology, and the polymer remains unknotted. The topological state of a loop
was characterized by $\varkappa$, the logarithm of the Alexander polynomial evaluated at $-1.1$
\cite{gr-krem2,knots-kardar,knots-kardar2}. To  ensure equilibration,  we estimated the scaling of
the equilibration time with $N$ for $N\le 32\,000$, extrapolated it to large $N$, and ran
simulations of longer chains, $N=108\,000$ and $256\,000$, to exceed the estimated equilibration
time. We also made sure that chains with topological constraints remain completely unknotted
through the simulations, while polymers with relaxed topological constraints become highly
entangled.

\begin{figure}[htbp]
\centerline{\includegraphics[width=12cm]{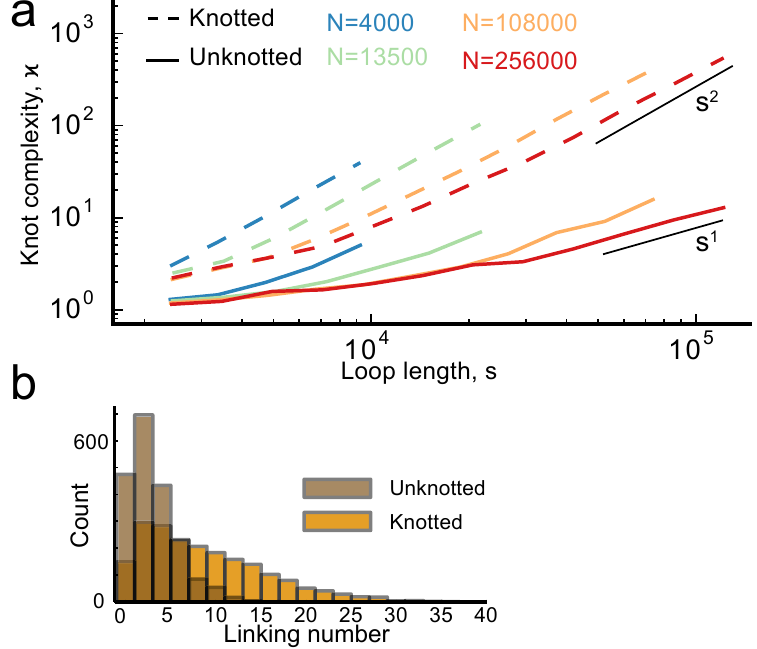}}
\caption{Topological properties of polymer loops in the knotted and unknotted globules.
(a) Knot complexity of polymer loops as a function of their length, $s$, for chains of different
length $N$ (shown by colors) in  knotted (dashed) and unknotted (solid) globules. (b) Distribution
of the linking numbers for non-overlapping loops of length $s=9000$ to $11\,000$ in $32\,000$-long
globules.}
\label{fig:13}
\end{figure}

To understand the role of topological correlations, we asked how the topological state of the whole
polymer influences the topological properties of its subchains. Because a topological state can be
rigorously defined only for a closed contour, we focused our analysis on loops, i.e. subchains with
two ends occupying neighboring lattice sites. The \fig{fig:13}a presents the average knot
complexity $\langle \varkappa(s) \rangle$ for loops of length $s$  for both types of globules. We
found that loops of the knotted globule were highly knotted, with the knot complexity rising
sharply with $s$. Loops of the unknotted globule, on the contrary, were weakly knotted, and their
complexity increased slowly with $s$.

\begin{figure}[htbp]
\centerline{\includegraphics[width=12cm]{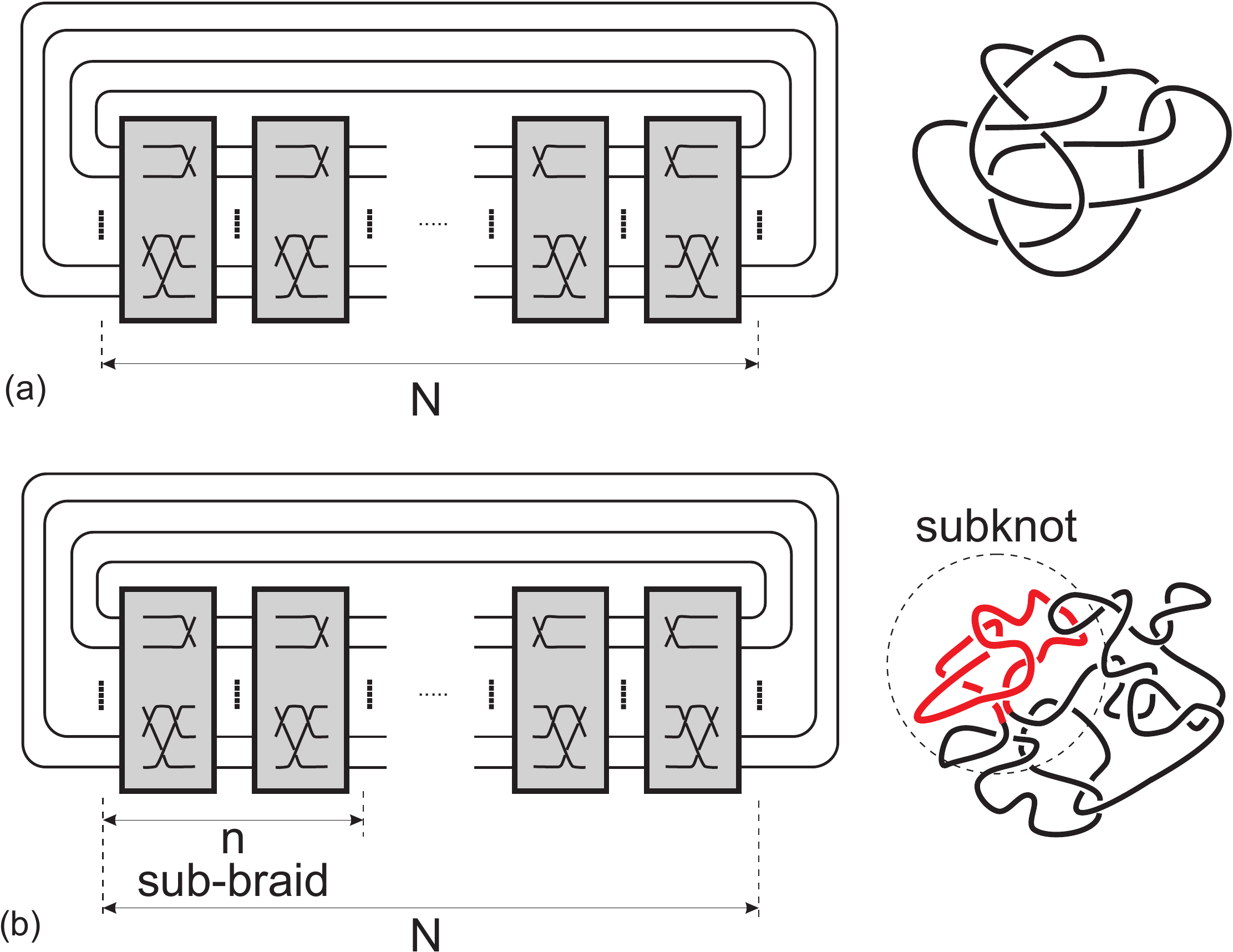}}
\caption{Schematic representation of knots by braids: a) unconditional random distribution of black
boxes produces a very complex knot; b) conditional distribution implies the whole knot to be
trivial, which imposed strong constraints on complexity of any subpart of the braid.}
\label{fig:14}
\end{figure}

This striking difference in the topological states of loops for globally knotted and unknotted
chains is a manifestation of the general statistical behavior of so-called matrix--valued Brownian
Bridges (BB) \cite{jktr}. The knot complexity $\varkappa$ of  loops in the topologically
unconstrained globule is expected to grow as $\varkappa(s)\sim s^2$. In contrast, due to the global
topological constraint imposed  on the unknotted globule, the knot complexity of its loops grows
slower, $\varkappa(s) \sim s$, which follows from the statistical behavior of BB in spaces of
constant negative curvature (see the Fig. \fig{fig:14}a, and \cite{vasne,jktr} for details).

Another topological property of loops of a globule is the degree of concatenation between the
loops. We computed the linking number for pairs of non-overlapping loops in the knotted and
crumpled (unknotted) globules, and found that loops in the unknotted globule are much less
concatenated than loops in the knotted globule. Taken together, these results show that the
topological state of the whole (parent) chain propagates to the daughter loops. While loops of the
unknotted globule are still slightly linked and knotted, their degree of entanglement is much lower
than for the loops in the topologically relaxed knotted globule.

Our topological problem to determine the complexity of a subloop in a globally trivial collapsed
polymer chain allows natural interpretation in terms of Brownian bridges. Suppose the following
imaginative experiment. Consider the phase space $\Omega$ of all topological states of densely
packed knots on the lattice. Select from $\Omega$ the subset $\omega$ of trivial knots. To simplify
the setting, consider a knot represented by a braid, as shown in the \fig{fig:14}, where the braid
is depicted by a sequence of uncorrelated "black boxes" (each black box contains some number of up-
and down-crossings. If crossings in all black boxes are identically and uniformly distributed, then
the boxes are statistically similar. Cut a part of each braid in the subset $\omega$, close open
tails and investigate the topological properties of resulting knots. Just such situation has been
qualitatively studied in \cite{vasne,jktr}. The crumpled globule hypothesis states the following:
if the whole densely packed lattice knot is trivial, then the topological state of each of its
"daughter" knot is almost trivial. We have shown that the computation of the knot complexity in the
braid representation depicted in the \fig{fig:14} can be interpreted as the computation of the
highest eigenvalue of the product of noncommutative matrices designated by black boxes.

To proceed, consider first the typical (unconditional) complexity of a knot represented by a
sequence of $N$ independent black boxes. This question is similar to the growth of the logarithm of
the largest eigenvalue, $\Lambda$, of the product of $N$ independent identically distributed
noncommutative random matrices. According to the F\"urstenberg theorem
\cite{furstenberg1963noncommuting}, in the limit $N\gg 1$ one has
\be
\ln \Lambda(N) \sim \gamma_1 N,
\label{eq:uncond}
\ee
where $\gamma_1={\rm const}$ is the so-called Lyapunov exponent (compare to \eq{eq:fuerst}). Being
rephrased for knots, this result means that the average knot complexity, $\varkappa$, understood as
a minimal number of crossings, $M$, necessary to represent a given knot by the compact knot
diagram, extensively grows with $M$, i.e. $\varkappa \sim M$. In the ordinary globule, for
subchains of length $N^{2/3} < s < N$, the typical number of crossing, $M$, on the knot diagram
grows as $M\sim s^2$, leading to the scaling behavior for the knot complexity $\varkappa$:
\be
\varkappa \sim s^2
\label{eq:vark1}
\ee
This is perfectly consistent with the well known fact: the probability of spontaneous unknotting of
a polymer with open ends in a globular phase is exponentially small. Following the standard scheme
\cite{volog, knots-kardar,knots-kardar2}, we characterize the knot complexity, $\varkappa$, by the
logarithm of the Alexander polynomial, $\ln [{\rm Al}(t=-1.1) {\rm Al}(t=-1/1.1)]$, i.e. we set
$\varkappa=\ln [{\rm Al}(t=-1.1)  {\rm Al}(t=-1/1.1)]$. As it seen from \fig{fig:13}, the
conjectured dependence $\ln {\rm Al}(t=-1.1)\sim s^2$ is perfectly satisfied for ordinary (knotted)
globule.

Consider now the \emph{conditional} distribution on the products of identically distributed black
boxes. We demand the product of matrices represented by black boxes to be a unit matrix
(topologically trivial). The question of interest concerns the typical behavior of $\ln
\Lambda^*(M,N)$, where $\Lambda^*(M,N)$ is the largest eigenvalue of the sub-chain of first $M$
matrices in the chain of $N$ ones. The answer to this question is known \cite{nesin2}: if $n=cN$
($0<c<1$ and $N\gg 1$), then
\be
\ln \lambda^*(n=cN,N) \sim \sqrt{c(1-c)} \sqrt{N} = \gamma_2(c) \sqrt{N}
\label{eq:cond}
\ee
where $\gamma_2(c)$ absorbs all constants independent on $N$. Translated to the knot language, the
condition for a product of $N$ matrices to be completely reducible, means that the "parent" knot is
trivial. Under this condition we are interested in the typical complexity $\varkappa^*$ of any
"daughter" sub-knot represented by first $n=cN$ black boxes.

Applying the \eq{eq:cond} to the knot diagram of the unknotted globule, we conclude that the
typical conditional complexity, $\varkappa^*$ expressed in the minimal number of crossings of any
finite sub-chain of a trivial parent knot, grows as
\be
\varkappa^*\sim \sqrt{s^2} \sim s
\label{eq:vark2}
\ee
with the subchain size, $s$. Comparing \eq{eq:vark2} and \eq{eq:vark1}, we conclude that subchains
of length $s$ in the trivial knot are much less entangled/knotted than subchains of same lengths in
the unconditional structure, i.e. when the constraint for a parent knot to be trivial is relaxed.
Indeed, this result is perfectly supported by \fig{fig:13} which show linear grows of
$\tilde{\varkappa} = \ln [{\rm Al}(t=-1.1)\,{\rm Al}(t=-1/1.1)]$ with $s$ for the unknotted
globule, while quadratic grows for the knotted globule.

\section{Conclusion}

\subsection{The King is dead, long live The King!}

The very concept of the crumpled (fractal) globule as of possible thermodynamic equilibrium state
of an unknotted ring polymer confined in a small volume, appeared in 1988 in a joint work by A.
Grosberg, S. Nechaev and E. Shakhnovich \cite{gns}. Soon after, in 1993, A. Grosberg, Y. Rabin, S.
Havlin, and A. Neer published a paper where they proposed the crumpled globule model to be a
possible condensed state of DNA packing in a chromosome \cite{gr-rab}. Then, over decades, the
interest to the crumpled globule was moderate: it was considered as an interesting, though
sophisticated artificial exercise. The attempts to find the crumpled structure in direct numeric
simulations, or in real experiments on proteins or DNAs were not too convincing.

Still, few interesting exceptions, which fuelled some discussions around the crumpled globule,
should be mentioned:

\begin{itemize}

\item the observation of the two-stage dynamics of collapse of the macromolecule after abrupt
changing of the solvent quality, found in light scattering experiments by B. Chu and Q. Ying
\cite{chu};

\item the experiments on compatibility enhancement in mixtures of ring and linear chains \cite{mck},
the construction of the quantitative theory of a collapse of $N$--isopropilacrylamide gel in a poor
water \cite{frac};

\item the experiments on superelasticity of polymer gels prepared in diluted solutions
\cite{urayama};

\item the indications of observation of the crumpled globule in numerical simulations
\cite{shakh,borovin}.

\end{itemize}

The breakthrough in the interest to the crumpled globule happen after the brilliant experimental
work of the MIT-Harvard team in 2009 \cite{mirny}. Immediately after, the concept of crumpled
(fractal) globule became the candidate for the new paradigm explaining many realistic features the
DNA packing and functioning in a human genome.

Analysis of chromatin folding in human genome based on a genome-wide chromosome conformation
capture method (Hi-C) \cite{3c,mirny} provides a comprehensive information on spatial contacts
between genomic parts and imposes essential restrictions on available 3D genome structures. The
experimental Hi-C maps obtained for various organisms and tissues \cite{mirny,dixon,drosoph,mouse,
sofueva,dekker_NRG,bacteria} display very rich structure in a broad interval of scales. The
researchers pay attention to the average contact probability, ${\cal P}(s)$, between two units of
genome separated by a genomic distance, $s$, which decays in typical Hi-C maps approximately as
${\cal P}(s)\sim 1/s$ (see \cite{mirny}).

The crumpled globule is a state of a polymer chain which in a wide range of scales is self-similar
and almost unknotted, forming a fractal space-filling-like structure. Both these properties,
self-similarity and absence of knots, are essential for genome folding: fractal organization makes
genome tightly packed in a broad range of scales, while the lack of knots ensures easy and
independent opening and closing of genomic domains, necessary for transcription
\cite{gr-rab,leonid_review}. In a three-dimensional space such a tight packing results in a
\emph{space-filling} with the fractal dimension $D_f=D=3$. The Hi-C contact probability, $P_{i,j}$,
between two genomic units, $i$ and $j$ in a $N$-unit chain, depends on a combination of structural
and energetic factors. Simple mean-field arguments (see, for example, \cite{mirny}) demonstrate
that in a fractal globule with $D_f=3$ the \emph{average} contact probability, ${\cal P}(s) =
(N-s)^{-1} \sum_{i=0}^{N-s} P_{i,i+s}$, between two units separated by the genomic distance
$s=|i-j|$, decays as ${\cal P}(s) \sim s^{-1}$. It should be noted that recent numeric simulations
\cite{gr-krem2, nech-mirny}, and more sophisticated arguments beyond the mean-field approximation
\cite{grosberg2012review, grosb_softmatter}, point out that the contact probability decays as
${\cal P}(s) \sim s^{-\gamma}$ with $\gamma\simeq 1.05 - 1.09$.

Despite the crumpled globule is our "favorite child", I should clearly state that it does not
explain exhaustively all details of the chromatin folding and definitely should be combined with
more refined models and concepts. Theoretical models of chromatin packing in the nucleus, which can
possibly explain the observed behavior of intra-chromosome Hi-C contact maps, split roughly into
two groups. The first group of works relies on specific interactions within the chromatin, like
loop or bridge formation, \cite{loops1,loops2,loops3, loops4,loops5,pnas,longowskiDif} and these
authors do not believe in crumpled globule, while the second group aims to explain the chromatin
structure in terms of large-scale topological interactions \cite{gr-rab,rosa_everaers,mirny,
leonid_review,grosberg2012review, rosa_everaers2, grosb_softmatter, nech-mirny, onefifth} based on
the crumpled model of the polymer globule \cite{gns}. For example, in \cite{nazarov} we combined
the assumption that chromatin can be considered as a heteropolymer chain with a quenched primary
sequence \cite{filion}, with the general hierarchical fractal globule folding mechanism. With this
conjecture we were able to reproduce the large-scale chromosome compartmentalization, not assumed
explicitly from the very beginning. To show the compatibility of the hierarchical folding of a
crumpled globule with the fine structure of experimentally observed Hi-C maps, we suggested in
\cite{nazarov} a simple toy model based on the crumpled globule folding principles together with
account of quenched disorder in primary sequence.

To summarize, my feeling of the current state of the art called "the crumpled globule" is
formulated in the title of this Section. All together we have come a long way from the
nonperturbative description of topological constraints in collapsed polymer phase to real
biophysical applications, we have understood on this way constructive connection of statistics of
polymer entanglements with Brownian bridges in the non-Euclidean geometry, we have got new results
for statistics of random braids, and finally we have explained some features of DNA packing in
chromosomes. Keeping our eyes open, we clearly see that the appearance of new experimental results
demonstrates that our initial topological arguments were too crude and too naive. However they gave
birth of new understanding of the role of topology in genomics and have led to new ideas and
methods which constitute the modern STATISTICAL TOPOLOGY OF POLYMERS.

\subsection{Where to go?}

I think we are only at the beginning of a highway, where the statistics of random walks is
interwined with the geometric group theory, algebraic topology and integrable systems in
mathematical physics, as well as has various incarnations in physics dealing with the crumpled
globule concept. Let me name some but a few such directions.

I would like to mention random walks on braid groups, grows of random heaps, viewed as random
sequential ballistic deposition (random "Tetris game"). Introducing the concept of the "locally
free group" as an approximant of the braid group, one can solve exactly the word problem in the
locally free group and obtain analytically the bilateral estimates (from above and from below) for
the growth of the volume of the braid group $B_n$ for arbitrary $n$ \cite{36,45,93}. Interesting
new results beyond the mean-field approximation for entanglement of threads in random braids have
been obtained recently in \cite{bunin}.

Sequential ballistic deposition (BD) with the next-nearest-neighbor interactions in a $N$-column
box can viewed as a time-ordered product of $N\times N$-matrices consisting of a single
$sl_2$-block which has a random position along the diagonal. One can interpret the uniform BD
growth as the diffusion in the symmetric space $H_N$. In particular, the distribution of the
maximal height of a growing heap can be connected to the distribution of the maximal distance for
the diffusion process in $H_N$, where the coordinates of $H_N$ can be interpreted as the
coordinates of particles of the one-dimensional Toda chain. The group-theoretic structure of the
system and links to some random matrix models was discussed in \cite{82}.

As concerns the impact of crumpled globule concept in physics, we have demonstrated that folding
and unfolding of a crumpled polymer globule can be viewed as a cascade of equilibrium phase
transitions in a hierarchical system, similar to the Dyson hierarchical spin model. Studying the
relaxation properties of the elastic network of contacts in a crumpled globule, we showed that the
dynamic properties of hierarchically folded polymer chains in globular phase are similar to those
of natural molecular machines (like myosin, for example). We discuss the potential ways of
implementations of such artificial molecular machine in computer and real experiments, paying
attention to the conditions necessary for stabilization of crumples under the fractal globule
formation in the polymer chain collapse \cite{machine} (see also the work on coalescence of
crumples in polymer collapse \cite{kardar-bunin}).

The ability of crumpled globule to act as molecular machine definitely provokes a new angle on an
eternal problem of the origin of life, related to overcoming the "error threshold" in producing and
selecting complex molecular structures during the prebiotic evolution \cite{avet1}. This permits us
to put forward a conjecture about a possibility for the crumpled globule to be a sort of the
"primary molecular machine" naturally formed under the prebiotic conditions. The primary crumpled
globule molecular machine (CGMM) made by polymers (not necessarily of biological nature), could
perform some specific functions typical for true biological molecular machines. The diversity of
CGMM may be concerned mainly with the attracting manifold, in which the CGMM action is performed.
This allows for functional variability without altering the structural archetype. Certainly, the
idea that crumpled globule could be the prebiotic molecular machine needs experimental
verification. However, the results \cite{machine,avet2} provide a rather optimistic view on the
evolutionary scenarios in which the primary molecular machines, themselves, are taken out of the
bio-molecular context. In this paradigm, the beginning of biological evolution is associated with
the spontaneous appearance of complex functional systems of primary "artificial" CGMM capable of
performing collective reproduction and autonomous behavior, which then are replaced in evolution by
more effective biomolecular systems. On this optimistic note I would like to end the story.

\bigskip

These notes are based on several lectures at the SERC School on Topology and Condensed Matter
Physics (organized in 2015 by RKM Vivekananda University at S.N Bose National Center for Basic
Sciences, Calcutta, India). I would thank Somen Bhattacharjee for kind invitation, opportunity to
explore some wonderful places in India and strong push to arrange lectures as a written text. The
topics discussed above summarize the subjects of millions conversations over many years with my
friends and colleagues, Alexander Grosberg, Anatoly Vershik, Vladik Avetisov, Leonid Mirny, Michael
Tamm. Particularly I would like to thank Maxim Frank-Kamenetskii and Alexander Vologodsky, whose
nice review in 1981 fuelled my interest to polymer topology, and to Alexey Khokhlov with whom we
got first results beyond the Abelian theory of polymer entanglements in 1985. The importance of
conditional Brownian bridge concept in statistics of non-commutative random walks was recognized in
joint work with Yakov Sinai in 1991. Especially I would like to highlight the role of Alexander
Grosberg whose ironic and deep comments and ideas tease and support me for more than a quarter of
century.

\end{document}